\documentclass[longauth]{aa}

\usepackage{graphicx}
\usepackage{txfonts}
\usepackage{gensymb}
\usepackage[colorlinks=true, linkcolor=blue, citecolor=blue, draft=False]{hyperref}
\usepackage{placeins}

\begin{document}

\title{\textit{JWST} CEERS probes the role of stellar mass and morphology in obscuring galaxies}

\author{
Carlos G{\'o}mez-Guijarro
\inst{1}
\and
Benjamin Magnelli
\inst{1}
\and
David Elbaz
\inst{1}
\and
Stijn Wuyts
\inst{2}
\and
Emanuele Daddi
\inst{1}
\and
Aur{\'e}lien Le Bail
\inst{1}
\and
Mauro Giavalisco
\inst{3}
\and
Mark Dickinson
\inst{4}
\and
Pablo G. P\'erez-Gonz\'alez
\inst{5}
\and
Pablo Arrabal Haro
\inst{4}
\and
Micaela B. Bagley
\inst{6}
\and
Laura Bisigello
\inst{7,8}
\and
V\'eronique Buat
\inst{9}
\and
Denis Burgarella
\inst{9}
\and
Antonello Calabr{\`o}
\inst{10}
\and
Caitlin M. Casey
\inst{6}
\and
Yingjie Cheng
\inst{3}
\and
Laure Ciesla
\inst{9}
\and
Avishai Dekel
\inst{11}
\and
Henry C. Ferguson
\inst{12}
\and
Steven L. Finkelstein
\inst{6}
\and
Maximilien Franco
\inst{6}
\and
Norman A. Grogin
\inst{12}
\and
Benne W. Holwerda
\inst{13}
\and
Shuowen Jin
\inst{14,15,16}
\thanks{Marie Curie Fellow}
\and
Jeyhan S. Kartaltepe
\inst{17}
\and
Anton M. Koekemoer
\inst{12}
\and
Vasily Kokorev
\inst{18}
\and
Arianna S. Long
\inst{6}
\thanks{NASA Hubble Fellow}
\and
Ray A. Lucas
\inst{12}
\and
Georgios E. Magdis
\inst{14,15,16}
\and
Casey Papovich
\inst{19,20}
\and
Nor Pirzkal
\inst{21}
\and
Lise-Marie Seill\'e
\inst{9}
\and
Sandro Tacchella
\inst{22,23}
\and
Maxime Tarrasse
\inst{1}
\and
Francesco Valentino
\inst{24,14}
\and
Alexander de la Vega
\inst{25}
\and
Stephen M. Wilkins
\inst{26,27}
\and
Mengyuan Xiao
\inst{28}
\and
{L. Y. Aaron} {Yung}
\inst{29}
\thanks{NASA Postdoctoral Fellow}
}

\institute{
 Universit{\'e} Paris-Saclay, Universit{\'e} Paris Cit{\'e}, CEA, CNRS, AIM, 91191, Gif-sur-Yvette, France
\email{carlos.gomezguijarro@cea.fr}
\and
Department of Physics, University of Bath, Claverton Down, Bath BA2 7AY, UK
\and
University of Massachusetts Amherst, 710 North Pleasant Street, Amherst, MA 01003-9305, USA
\and
NSF's National Optical-Infrared Astronomy Research Laboratory, 950 N. Cherry Ave., Tucson, AZ 85719, USA
\and
Centro de Astrobiolog\'{\i}a (CAB), CSIC-INTA, Ctra. de Ajalvir km 4, Torrej\'on de Ardoz, E-28850, Madrid, Spain
\and
Department of Astronomy, The University of Texas at Austin, Austin, TX, USA
\and
INAF--Osservatorio Astronomico di Padova, Vicolo dell'Osservatorio 5, I-35122, Padova, Italy
\and
Dipartimento di Fisica e Astronomia "G.Galilei", Universit\'a di Padova, Via Marzolo 8, I-35131 Padova, Italy
\and
Aix Marseille Univ, CNRS, CNES, LAM Marseille, France
\and
INAF - Osservatorio Astronomico di Roma, via di Frascati 33, 00078 Monte Porzio Catone, Italy
\and
Racah Institute of Physics, The Hebrew University of Jerusalem, Jerusalem 91904, Israel
\and
Space Telescope Science Institute, 3700 San Martin Drive, Baltimore, MD 21218, USA
\and
Physics \& Astronomy Department, University of Louisville, 40292 KY, Louisville, USA
\and
Cosmic Dawn Center (DAWN), Denmark
\and
DTUSpace, Technical University of Denmark, Elektrovej 327, 2800 Kgs. Lyngby, Denmark
\and
Niels Bohr Institute, University of Copenhagen, Jagtvej 128, 2200 Copenhagen, Denmark
\and
Laboratory for Multiwavelength Astrophysics, School of Physics and Astronomy, Rochester Institute of Technology, 84 Lomb Memorial Drive, Rochester, NY 14623, USA
\and
Kapteyn Astronomical Institute, University of Groningen, P.O. Box 800, 9700AV Groningen, The Netherlands
\and
Department of Physics and Astronomy, Texas A\&M University, College Station, TX, 77843-4242 USA
\and
George P.\ and Cynthia Woods Mitchell Institute for Fundamental Physics and Astronomy, Texas A\&M University, College Station, TX, 77843-4242 USA
\and
ESA/AURA Space Telescope Science Institute
\and
Kavli Institute for Cosmology, University of Cambridge, Madingley Road, Cambridge, CB3 0HA, UK
\and
Cavendish Laboratory, University of Cambridge, 19 JJ Thomson Avenue, Cambridge, CB3 0HE, UK
\and
European Southern Observatory, Karl-Schwarzschild-Str. 2, D-85748 Garching bei Munchen, Germany
\and
Department of Physics and Astronomy, University of California, 900 University Ave, Riverside, CA 92521, USA
\and
Astronomy Centre, University of Sussex, Falmer, Brighton BN1 9QH, UK
\and
Institute of Space Sciences and Astronomy, University of Malta, Msida MSD 2080, Malta
\and
Department of Astronomy, University of Geneva, Chemin Pegasi 51, 1290 Versoix, Switzerland
\and
Astrophysics Science Division, NASA Goddard Space Flight Center, 8800 Greenbelt Rd, Greenbelt, MD 20771, USA
}

\date{}

\abstract
{In recent years, observations have uncovered a population of massive galaxies that are invisible or very faint in deep optical/near-infrared (near-IR) surveys but brighter at longer wavelengths. However, the nature of these optically dark or faint galaxies (OFGs; one of several names given to these objects) is highly uncertain. In this work, we investigate the drivers of dust attenuation in the \textit{JWST} era. In particular, we study the role of stellar mass, size, and orientation in obscuring star-forming galaxies (SFGs) at $3 < z < 7.5$, focusing on the question of why OFGs and similar galaxies are so faint at optical/near-IR wavelengths. We find that stellar mass is the primary proxy for dust attenuation, among the properties studied. Effective radius and axis ratio do not show a clear link with dust attenuation, with the effect of orientation being close to random. However, there is a subset of highly dust attenuated ($A_V > 1$, typically) SFGs, of which OFGs are a specific case. For this subset, we find that the key distinctive feature is their compact size (for massive systems with $\log (M_{*}/M_{\odot}) > 10$); OFGs exhibit a 30\% smaller effective radius than the average SFG at the same stellar mass and redshift. On the contrary, OFGs do not exhibit a preference for low axis ratios (i.e., edge-on disks). The results in this work show that stellar mass is the primary proxy for dust attenuation and compact stellar light profiles behind the thick dust columns obscuring typical massive SFGs.}

\keywords{galaxies: evolution -- galaxies: high-redshift -- galaxies: photometry -- galaxies: star formation -- galaxies: structure -- infrared: galaxies}

\titlerunning{\textit{JWST} CEERS probes the role of stellar mass and morphology in obscuring galaxies}
\authorrunning{C. G\'omez-Guijarro et al.}

\maketitle

\section{Introduction} \label{sec:intro}

Until \textit{JWST} came online, our understanding of the early ($z > 3$) cosmic history of galaxies was mainly based on samples selected from their rest-frame ultraviolet (UV) light \citep[see][for a review]{madau14}. However, this view is limited, as classical color selections to find distant star-forming galaxies (SFGs), such as the Lyman break galaxies \citep[LBGs; e.g.,][]{steidel96,giavalisco04,bouwens12}, are effective at identifying blue galaxies with low dust attenuation, but are generally biased against redder dusty SFGs \citep[see][for a review]{casey14,hodge20}.

In recent years, Spitzer Space Telescope (\textit{Spitzer}) and Atacama Large Millimeter/submillimeter Array (ALMA) observations from numerous studies have uncovered a population of massive dusty SFGs completely missed in deep optical/near-infrared (optical/near-IR) surveys, but bright at longer far-infrared/millimeter (far-IR/mm) wavelengths \citep[see e.g.,][so-called optically dark/faint galaxies (OFGs), or, alternatively, optically invisible galaxies, optical/near-IR-dark, \textit{HST}-dark, among others]{simpson14,wang16,wang19,franco18,alcaldepampliega19,yamaguchi19,williams19,romano20,toba20,umehata20,zhou20,gruppioni20,talia21,smail21,fudamoto21,manning22,gomezguijarro22a,shu22,enia22,xiao23}. These dusty SFGs are different from the rare extreme dusty starbursts \citep[e.g.,][]{walter12,riechers13,marrone18}, as they are more numerous, with number densities two orders of magnitude greater, and are characterized by milder star formation rates \citep[SFRs; e.g.,][]{wang19} typical of galaxies in the main sequence of SFGs \citep[MS; e.g.,][]{brinchmann04,daddi07,elbaz07,noeske07,whitaker12,speagle14,schreiber15}. These galaxies are believed to be a dominant contributor to the massive ($M_{*} > 10^{10.3}$\,$M_{\odot}$) end of the SFR density of the Universe at $z \sim 3$--6 \citep[e.g.,][]{wang19,xiao23,barrufet23}. However, the dearth of available secure spectroscopic redshifts and determinations of their stellar mass is a hinderance to the interpretation of these sources among the SFG population. It is still unclear as to how many owe their faintness at optical/near-IR wavelengths to a dusty or a quiescent nature. In addition, being so faint or completely dark to the Hubble Space Telescope (\textit{HST}) has rendered the study and understanding of the role of their stellar morphologies impossible.

The \textit{JWST} recently began to be used to investigate this galaxy population, unveiling the stellar light of galaxies previously undetected by \textit{HST} (\textit{HST}-dark). In a general sense, it should be noted that there is no consensus definition of this galaxy population and the properties of its members heavily depend on the specific magnitude and color cuts \citep[e.g.,][]{xiao23}. Among these \textit{JWST} studies, \citet{barrufet23}, based on a red ($F160W$ - $F444W$) color selection, reported their relatively high-redshift, massive, MS-like SFRs, a dusty nature ($z \sim 2$--8; $M_{*} > 10^{10}$\,$M_{\odot}$; $A_V \sim 2$\,mag), and their important contribution to the cosmic SFR density at $z \sim 6$. \citet{nelson23} measured a prevalence of low axis ratios and linked their elusiveness to a dusty disk-like and edge-on nature. \citet{perezgonzalez23}, based on a red ($F150W$ - $F356W$) color selection, described this galaxy population as threefold, including a majority ($\sim 70$\%) of dusty SFGs at $z = 2$--6, with $M_{*} = 10^{9-10}$\,$M_{\odot}$, $\sim 20$\% quiescent galaxies (QGs) with $M_{*} > 10^{10}$\,$M_{\odot}$ at $z = 3$--4, and $\sim 10$\% young starbursts at $z = 6$--7 with $M_{*} \sim 10^{9.5}$\,$M_{\odot}$. \citet{rodighiero23} reported evolved $z = 8$--13 massive $M_{*} = 10^{9-10}$\,$M_{\odot}$ galaxies with high dust attenuation ($A_V > 5$\,mag). \citet{kokorev23} presented a spatially resolved study of a $z \sim 2.5$ massive ($M_{*} \sim 10^{11.3}$\,$M_{\odot}$) dark galaxy, finding very high levels of dust attenuation ($A_V \sim 4$\,mag) all across its stellar extent.

In this work, we investigate the drivers of dust attenuation in the \textit{JWST} era, focusing on understanding the nature of obscured galaxies to unveil the reason behind their elusiveness in the pre-\textit{JWST} era. We study the roles of stellar mass and morphology as proxies for dust attenuation. The layout of the paper is as follows. The data used in this work are described in Sect.~\ref{sec:data}. In Sect.~\ref{sec:catalog}, we present the catalog and photometry, along with the derivation of stellar-based properties, including morphological measurements. Section~\ref{sec:sample} describes the sample selection, classification into LBGs and OFGs, and evaluation of the star-forming or quiescent nature of OFGs. In Sect.~\ref{sec:prop}, we characterize the stellar-based properties, including redshift, stellar mass, dust attenuation, SFR, and morphological measurements. We investigate the relevance of the stellar mass and morphology in driving dust attenuation in Sect.~\ref{sec:av_drivers}. In Sect.~\ref{sec:discussion}, we discuss and interpret our results. We summarize our main findings and conclusions in Sect.~\ref{sec:summary}.

Throughout this work, we adopt a concordance cosmology $[\Omega_\Lambda,\Omega_M,h]=[0.7,0.3,0.7]$ and a Chabrier initial mass function (IMF) \citep{chabrier03}. When magnitudes are quoted, they are in the AB system \citep{oke74}.

\section{Data} \label{sec:data}

\subsection{\textit{JWST} data} \label{subsec:jwst_data}

In this work, we used \textit{JWST} data from the Cosmic Evolution Early Released Science (CEERS) survey, an early release science program in the Extended Groth Strip (EGS) field \citep{finkelstein22}. The CEERS data include \textit{JWST}/NIRCam imaging over ten pointings, the first four (CEERS1, CEERS2, CEERS3, CEERS6) observed in June 2022 and the remaining six (CEERS4, CEERS5, CEERS7-10) observed in December 2022, together covering a total area of 97\,arcmin$^2$. This dataset comprises observations with seven NIRCam filters $F115W$, $F150W$, $F200W$, $F277W$, $F356W$, $F410M$, and $F444W$, reaching an average 5$\sigma$ point source depth of 29.15, 29.00, 29.17, 29.19, 29.17, 28.38, and 28.58\,mag, respectively.

For details about the observations and data reduction process, we refer the reader to \citet{bagley23}. Briefly, the June pointings were processed using the \textit{JWST} Calibration Pipeline version 1.7.2 and CRDS pmap 0989, while the December pointings were processed using the pipeline version 1.8.5 and CRDS pmap 1023. For both June and December pointings, the reduction steps were the same, with the raw images passing through all pipeline stages with some additional steps, including $1/f$ noise subtraction and wisp removal. The mosaics were aligned to the Gaia-EDR3 astrometry \citep{gaia21}, mapped into a pixel scale of 0\farcs03/pixel, and background subtracted.

\subsection{Ancillary data} \label{subsec:anc_data}

In order to complement shorter wavelengths, we used \textit{HST} imaging in the EGS field from its public release reprocessed by the CEERS team, matching the astrometry and pixel scale of the \textit{JWST} data \citep[see][]{koekemoer11}. This dataset comprises observations with six filters: ACS/WFC $F606W$, $F814W$, WFC3/IR $F105W$, $F125W$, $F140W$, and $F160W$, reaching an average 5$\sigma$ point source depth of 28.62, 28.30, 27.11, 27.31, 26.67, and 27.37\,mag, respectively.

We also used the publicly available ground-based imaging from the Canada-France-Hawaii Telescope (CFHT)/MegaCam observations in the EGS field in $u^*$, $g'$, $r'$, $i'$, and $z'$-band, reaching average 5$\sigma$ point source depths of 27.1, 27.3, 27.2, 27.0, and 26.1\,mag, respectively. This dataset corresponds to the D3 field of the CFHT Legacy Survey (CFHTLS) \citep{gwyn12}.

\section{Catalog} \label{sec:catalog}

\subsection{Source detection and photometry} \label{subsec:se_phot}

For source detection, we followed a similar approach to that used in the CANDELS catalog in the EGS field \citep{stefanon17}. We employed \texttt{SExtractor} \citep{bertin96} using the $F444W$-band as a detection image. This band was found to be optimal for our science case, because our goal is to unveil galaxies that become increasingly faint at shorter wavelengths. The $F444W$-band comprises the longest wavelengths of the \textit{JWST}/NIRCam bands. We carried out the source detection in both the so-called cold and hot modes, that is, with tailored parameters optimized to detect from bright extended to faint compact sources. We also employed a RMS map (defined as $\rm{RMS} = 1 / \sqrt{\rm{WHT}}$, where the WHT map is a weight image giving the relative weight of the output pixels) scaled to the median value of the ERR map (data array containing resampled uncertainty estimates, given as standard deviation) as a weighting image for detection. The final merged catalog comprises all cold sources and the hot sources that are not part of a cold source. Discarded hot sources are those that fall within the Kron ellipse of a cold source.

In the 13 \textit{JWST} and \textit{HST} bands, we measured fluxes using \texttt{SExtractor} in dual image mode. In this mode, one image is used for detections ($F444W$-band in our case), while measurements are carried out on another image. Running \texttt{SExtractor} with various measurement images while keeping the same detection image, one ends up with a catalog with the same sources measured through different bands. Measurements were carried out on images PSF-matched to the angular resolution of the detection $F444W$-band image (0\farcs16 FWHM), which provides the coarsest angular resolution of the \textit{JWST}/NIRCam bands and a resolution that is comparable to those of the \textit{HST} WFC3 bands. PSF-matching was performed on \textit{JWST}/NIRCam and \textit{HST}/ACS images by convolving them with a convolution kernel created using the software \texttt{PyPHER} \citep{boucaud16}, from PSF images of the different bands constructed by stacking point sources that were selected in the mosaics using the software \texttt{PSFEx} \citep{bertin11}. Total flux measurements and uncertainties in both the cold and hot modes in the detection $F444W$-band image were calculated using Kron elliptical apertures (\texttt{FLUX\_AUTO} with \texttt{PHOT\_AUTOPARAMS = 2.5, 3.5}). For the remaining bands, \texttt{SExtractor} dual mode uses the segmentation map of the detection image to measure isophotal fluxes and uncertainties (\texttt{FLUX\_ISO}) that we corrected to total fluxes by multiplying them with the ratio $\rm{\texttt{FLUX\_AUTO}}/\rm{\texttt{FLUX\_ISO}}$ as measured in the detection $F444W$-band \citep[see][]{stefanon17}. \textit{HST}/WFC3 fluxes and uncertainties were scaled to account for the missing flux when comparing their PSF with the $F444W$-band PSF. An additional aperture correction optimized for point sources was added in all bands to account for the flux missed outside of the Kron elliptical apertures. For each source, we scaled their fluxes and uncertainties by the fraction of flux outside of the area covered by its Kron ellipse as measured in the $F444W$-band PSF.

In the five ground-based CFHT bands, we measured fluxes using aperture photometry on images PSF-matched to the angular resolution of the $u*$-band (0\farcs9 FWHM), which provides the coarsest angular resolution of the CFHT bands (i.e., the poorest seeing conditions at the moment of observation). We measured fluxes in 1\farcs2 diameter apertures, which provides an optimal trade-off between total flux retrieval and signal-to-noise ratio \citep[S/N; see also, e.g.,][]{straatman16}. Total flux measurements were then calculated using aperture corrections derived tracing the $u*$-band PSF growth curve to account for the flux losses outside the chosen aperture. 

Fluxes and uncertainties in all bands were corrected for Milky Way attenuation following a \citet{cardelli89} extinction function with an average $E(B-V) = 0.006$ in the EGS field \citep{schlafly11}.

\subsection{Redshift, stellar mass, and dust attenuation} \label{subsec:z_mstar_av}

We estimated photometric redshifts from spectral energy distribution (SED) fitting to the 18 bands in our catalog ($u^*$ to $F444W$) using the code \texttt{EAZy} \citep{brammer08} in its updated \texttt{Python} version \texttt{EAZy-py}. We employed the set of 13 Flexible Stellar Population Synthesis \citep[FSPS;][]{conroy10} templates \citep[\texttt{corr\_sfhz\_13}; see][for details]{kokorev22}. These templates cover redshift-dependent star formation histories (SFHs), which disfavor SFHs that start earlier than the age of the Universe at a given epoch, span a large range in ages and dust attenuation, and include emission lines. Additionally, the set was complemented with the best-fit template to the extreme emission line galaxy 4590 at $z = 8.5$ with a \textit{JWST}/NIRSpec spectrum \citep{carnall23} rescaled to match the normalization of the FSPS templates \citep[see e.g.,][for a similar approach]{kokorev22,gould23,valentino23}. \texttt{EAZy-py} finds the best linear combination of templates for the flux densities and uncertainties of each source. We searched for updated spectroscopic redshifts by cross-matching the source catalog ($r < 0\farcs5$) with the ancillary spectroscopic compilation in \citet{stefanon17} and the MOSDEF survey in the EGS field \citep{kriek15}. When a high-quality measurement was available, we substituted the photometric redshift by its spectroscopic value. In addition, we used the best-quality spectroscopic redshifts to derive zero-point corrections applied iteratively to the \texttt{EAZy-py} run to improve photometric redshifts ($N = 320, \sigma = 0.02$, corrections $< 10$\%).

The set of \texttt{EAZy} templates is best suited for redshift determination, but they are limited in terms of their physical parameters. Therefore, we estimated stellar masses and dust attenuation ($A_V$) using the SED fitting code \texttt{FAST++}\footnote{https://github.com/cschreib/fastpp}, which is an updated version of the SED fitting code \texttt{FAST} \citep{kriek09}, fixing the redshifts to the values previously obtained. The stellar population models were from \citet{bruzual03} (BC03), with delayed exponentially declining SFHs ($\tau = 100$\,Myr--30\,Gyr; age $= 50$\,Myr--age of the universe at a given redshift), a \citet{calzetti00} dust attenuation law ($A_{V} = 0.0$--5.0) ---which is common in similar studies and provides a benchmark for comparison--- and metallicity ($Z = 0.004$, 0.008, 0.02 (solar), 0.05).

\subsection{Star formation rate} \label{subsec:sfr}

We estimated SFR values from a ladder of SFR estimators \citep[see e.g.,][for a similar approach]{wuyts11,magnelli14,barro19}.

First, we checked whether there are counterparts in the mid-IR-to-mm bands in the "super-deblended" catalog in the EGS field (Le Bail et al., in prep). This catalog is built using a state-of-the-art de-blending methodology similar to that of similar catalogs in the GOODS-North \citep{liu18} and COSMOS \citep{jin18} fields. These mid-IR-to-mm bands include \textit{Spitzer}/MIPS/24\,$\mu$m, \textit{Herschel}/PACS/100, 160\,$\mu$m, \textit{Herschel}/SPIRE/250, 350, 500\,$\mu$m, JCMT/SCUBA2/450, 850\,$\mu$m, and ASTE/AzTEC 1.1\,mm. For galaxies with $\rm{S/N_{IR}} > 5$ (sum in quadrature of the S/N of all the bands beyond and not including \textit{Spitzer}/MIPS/24\,$\mu$m), we fit \citet{draineli07} dust emission templates with the correction from \citet{draine14} using the code \texttt{CIGALE} \citep{burgarella05,noll09,boquien19} to derive total IR luminosity estimates ($L_{\rm{IR}}$, 8--1000\,$\mu$m rest-frame). For galaxies with $\rm{S/N_{IR}} < 5$, but $\rm{S/N_{24\mu m}} > 5$, we renormalized IR templates from \citet{schreiber18} to the \textit{Spitzer}/MIPS/24\,$\mu$m fluxes, obtaining $L_{\rm{IR}}$. The total SFR accounts for the contribution of the obscured star formation probed in the IR ($\rm{SFR_{IR}}$) and the unobscured star formation probed in the UV ($\rm{SFR_{UV}}$), following the prescription of \citet{bell05} (for a \citet{chabrier03} IMF):

\begin{equation}
\label{eq:sfr}
\rm{SFR} = \rm{SFR_{IR}} + \rm{SFR_{UV}} = 1.09 \times 10^{-10} (\textit{L}_{\rm{IR}} + 2.2\textit{L}_{\rm{UV}})\,, 
\end{equation}

\noindent where $\textit{L}_{\rm{UV}} = 1.5 \nu l_{\nu,2800}$ is the total UV luminosity in the range 1260--3000\,\AA~ rest frame derived from the rest frame 2800\,\AA~ luminosity density ($l_{\nu,2800}$).

For galaxies with $\rm{S/N_{IR}} < 5$ and $\rm{S/N_{24\mu m}} < 5$, or for those without counterparts in the mid-IR-to-mm bands in the super-deblended catalog, the total SFR is given by the $\rm{SFR_{UV}}$ corrected from dust attenuation ($\rm{SFR_{UV,cor}}$). We calculated $\rm{SFR_{UV,cor}}$ following the same prescription above for the UV term, but with $\textit{L}_{\rm{UV,cor}} = 1.5 \nu l_{\nu,2800,\rm{cor}}$, and with  $l_{\nu,2800,\rm{cor}}$ being the rest-frame 2800\,\AA~ luminosity density corrected for dust attenuation ($A_{2800}$) using the \citet{calzetti00} law.

\subsection{Morphology} \label{subsec:morph}

In order to estimate morphological structural parameters, we used surface brightness profile fitting to the \textit{JWST}/NIRCam $F444W$ images by employing the code \texttt{statmorph} \citep{rodriguezgomez19}. The $F444W$-band is the reddest NIRCam filter available, probing $0.52 < \lambda_{\rm{rf}} < 1.1$\,$\mu$m in the redshift range of interest ($3 < z < 7.5$, see Sect.~\ref{sec:sample}) at $\sim 0\farcs16$ angular resolution (PSF FWHM). Therefore, the $F444W$-band probes optical-to-near-IR rest-frame morphologies, mitigating the morphological differences that can occur at shorter wavelengths, especially in the rest-frame UV \citep[see e.g.,][]{suess22,miller23}. \texttt{statmorph} performs 2D S\'ersic profile fitting and nonparametric morphological diagnostics of a given background-subtracted image and associated segmentation map with the sources of interest. The code includes a \textit{flag} for the fit quality, where $flag = 0$ indicates a good fit and $flag \geq 2$ indicates a problem with the measurements (e.g., artefacts, foreground stars, or incompletely masked secondary sources). Following the documentation, we restricted our analysis to $flag \leq 1$ and $flag\_sersic = 0$ when considering morphological measurements.

\section{Sample definitions} \label{sec:sample}

\subsection{General parent SFGs sample} \label{subsec:sfgs_sample}

In order to include new insight from \textit{JWST}/NIRCam into the rest-frame UV, we introduced a lower limit in the redshift for our study. We focused our work on the redshift range $3 < z < 7.5$. The lower redshift limit was imposed to cover the rest-frame UV ($\lambda_{\rm{rf}} < 4000$\,\AA) at all redshifts with at least two NIRCam bands ($F115W$ and $F150W$), while the higher redshift limit was imposed because of the scarcity of galaxies at higher redshifts. We also introduced a lower limit in the stellar mass of $\log (M_{*}/M_{\odot}) \geq 9.0$ to focus our work on intermediate-to-massive galaxies.

We classified this redshift- and stellar mass-limited sample into SFGs and QGs through a $UVJ$ classification. The $UVJ$ plane is a classical color diagram widely used to make this discrimination \citep[e.g.,][]{wuyts07,williams09,patel12}. We employed the definition by \citet{williams09}. The rest-frame colors were obtained from \texttt{FAST++} and the resulting best-fit SEDs. Although the $UVJ$ diagram has proven to be a useful tool for SFGs and QGs discrimination, it is not free of some plausible contamination. In order to mitigate contaminants, we checked that the galaxies classified as QGs by the $UVJ$ criteria are undetected ($< 2\sigma$) in mid-IR-to-mm bands (from \textit{Herschel}/PACS/100\,$\mu$m to ASTE/AzTEC 1.1\,mm) in the super-deblended catalog in the EGS field (Le Bail et al., in prep); otherwise we classified them as SFGs.

As a result, we defined a redshift- ($3 < z < 7.5$) and stellar mass-limited ($\log (M_{*}/M_{\odot}) \geq 9.0$) sample of $UVJ$-selected SFGs (parent SFGs sample), which serves as a benchmark for comparisons. In total, the parent SFG sample comprises 1490 galaxies.

\begin{figure}
\begin{center}
\includegraphics[width=\columnwidth]{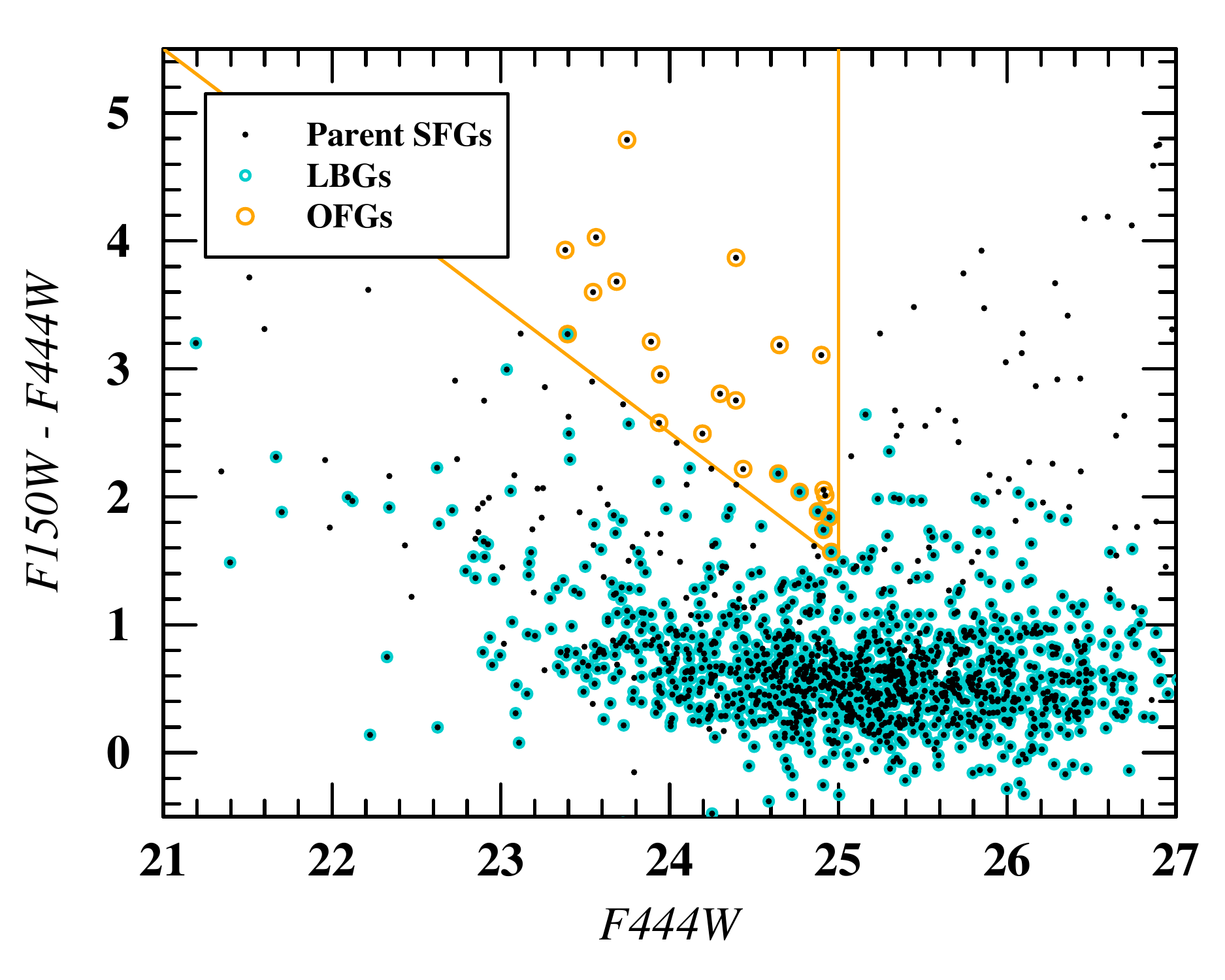}
\caption{($F150W - F444W$) color versus $F444W$ magnitude distribution for the parent SFG (black), LBG (cyan circles), and OFG (orange circles) samples. The OFG selection criteria is displayed with orange segments.}
\label{fig:col_mag}
\end{center}
\end{figure}

\subsection{Lyman break galaxies} \label{subsec:lbgs_sample}

The Lyman limit ($\lambda_{\rm{rf}} = 912$\,\AA) is a signature that allows the identification of high-redshift SFGs. SFGs exhibit a prominent break beyond their Lyman limit owing to absorption by intervening hydrogen. These galaxies are called LBGs. An LBG will appear very faint or invisible in all bands bluewards of its Lyman limit and prominent again redwards of its Lyman limit. The more redshifted the Lyman limit, the more distant the galaxy. Therefore, the Lyman break technique is an efficient way of selecting high-redshift SFGs \citep[e.g.,][]{steidel96,giavalisco04,bouwens12}. The technique was used up to $z \sim 10$ in the pre-\textit{JWST} era \citep[e.g.,][]{bouwens15}.

In this work, we label LBGs in the parent SFGs sample galaxies; these are selected by applying the LBG color criteria as defined by \citet{bouwens20}:

\begin{eqnarray*}
z\sim3: (U_{336}-B_{435}>1)\wedge(B_{435}-V_{606}<1.2)\wedge \\
(i_{775}-Y_{105}<0.7),
\end{eqnarray*}
\begin{eqnarray*}
z\sim4: (B_{\rm 435}-V_{\rm 606}>1)\wedge (i_{\rm 775}-J_{\rm 125}<1) \wedge \\
(B_{\rm 435}-V_{\rm 606} > 1.6(i_{\rm 775}-J_{\rm 125})+1),
\end{eqnarray*}
\begin{eqnarray*}
z\sim5: (V_{\rm 606}-i_{\rm 775}>1.2)\wedge (z_{\rm 850}-H_{\rm 160}<1.3) \wedge \\
(V_{\rm 606}-i_{\rm 775} > 0.8(z_{\rm 850}-H_{\rm 160})+1.2),
\end{eqnarray*}
\begin{eqnarray*}
z\sim6: (i_{\rm 775}-z_{\rm 850}>1.0)\wedge (Y_{\rm 105}-H_{\rm 160}<1.0) \wedge \\
(i_{\rm 775}-z_{\rm 850} > 0.777(Y_{\rm 105}-H_{\rm 160})+1.0),
\end{eqnarray*}
\begin{eqnarray*}
z\sim7: (z_{\rm 850}-Y_{\rm 105}>0.7)\wedge (J_{\rm 125}-H_{\rm 160}<0.45) \wedge \\
(z_{\rm 850}-Y_{\rm 105} > 0.8(J_{\rm 125}-H_{\rm 160})+0.7), 
\end{eqnarray*}

\noindent where $\wedge$ and $\vee$ represent the logical AND and
OR symbols, respectively. The bands involved in these LBG color criteria correspond to \textit{HST}. In order to properly mimic this selection in the absence of observations in the exact same bands, the magnitudes in the relevant filters were calculated by convolving their transmission curves with the resulting best-fit SEDs from \texttt{FAST++}. In addition, to avoid classifying galaxies based on model extrapolations of observationally unconstrained regions of the SED, we required at least two (> 5$\sigma$) detections in the rest-frame UV ($\lambda_{\rm{rf}} < 4000$\,\AA) of the SED to perform the classification. The LBGs sample comprises 1182 galaxies within the parent SFGs sample.

\subsection{Optically dark/faint galaxies} \label{subsec:ofgs_sample}

Optically dark/faint galaxies (OFGs) is a general term referring to the intrinsic faintness of these galaxies around optical/near-IR wavelengths \citep{gomezguijarro22a,xiao23}. In this sense, they are defined as galaxies undetected or very faint in all optical and near-IR bands up to and including the $H$-band in the deepest cosmological fields (typical $5\sigma$ point source depth $H > 27$\,mag) but bright at longer near-IR bands ($[3.6]$ or $[4.5]$-band). Their redshifts, stellar masses, dust attenuation, and star-forming or quiescent nature heavily depend on the specific magnitude and color cuts.

A typical selection consists in targeting sources that are $H$-band dropouts (with studies defining them below a given magnitude limit or as absent in catalogs) but bright at 3.6/4.5\,$\mu$m, or galaxies with very red colors (e.g., $H - 3.6/4.5 > 2$--4\,mag) \citep[see e.g.,][]{wang16,wang19,alcaldepampliega19,perezgonzalez23}. This selection typically yields $z > 3$ massive and dust obscured galaxies \citep[see also][for low-mass dusty SFG selection criteria]{bisigello23}. \citet{xiao23}, with the purpose of bridging more extreme optically dark galaxies with more common lower-mass, less-dust attenuated galaxies, such as LBGs, formally define OFGs based on the following criteria: (1) $H > 26.5$\,mag; (2) $[4.5] < 25$\,mag. The $H > 26.5$ cut selects both the extreme optically dark and intrinsically fainter galaxies with less dust attenuation. This cut also weeds out typical massive ($\log (M_{*}/M_{\odot}) > 10$) QGs. The $[4.5] < 25$\,mag cut selects not only massive galaxies, but also galaxies with intermediate stellar masses \citep[see][for details]{xiao23}.

In this work, we applied the magnitude cuts as defined by \citet{xiao23} using the new insight from the \textit{JWST}/NIRCam bands: (1) $F150W > 26.5$\,mag; (2) $F444W < 25$\,mag. This parent OFG sample comprises 35 galaxies. There are 25/35 (71\%) in the parent SFG sample and 3/35 (8.6\%) QGs based on the $UVJ$ classification. The remaining 7/35 (20\%) galaxies belong to a lower-redshift tail ($1.5 < z < 3$). This tail is left out of the parent SFG sample when we impose the lower redshift limit ($z > 3$) to cover the rest-frame UV ($\lambda_{\rm{rf}} < 4000$\,\AA) at all redshifts with at least two NIRCam bands ($F115W$ and $F150W$). Nevertheless, the OFG criteria appear to be an effective filter for relatively high-redshift ($3 < z < 7.5$), intermediate-to-massive ($M_{*} > 10^{9}$\,$M_{\odot}$), dusty ($A_V \gtrsim 1$) SFGs (25/35, 71\%), with little contamination from QGs (3/35, 8.6\%). Hereafter, we continued the analysis focusing on the OFGs sample that is part of the parent SFGs sample (see Sect.~\ref{sec:prop}, for details on their redshift, stellar mass, and dust attenuation distributions).

\begin{figure}
\begin{center}
\includegraphics[width=\columnwidth]{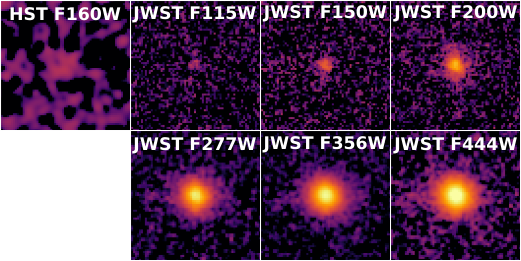}
\caption{Example OFG at $z_{\rm{phot}} = 3.52$, with $\log (M_{*}/M_{\odot}) = 10.39$ and $A_V = 2.16$. While the galaxy is not detected ($\rm{S/N} < 1$) in the \textit{HST}/WFC3 F160W image, it is detected ($\rm{S/N} > 5$) in all \textit{JWST}/NIRCam bands. Images are 2\arcsec$\times$2\arcsec with north up and east to the left.}
\label{fig:ofgs_example}
\end{center}
\end{figure}

Figure~\ref{fig:col_mag} shows the ($F150W - F444W$) color versus $F444W$ magnitude distribution, including the parent SFGs, LBGs, and OFGs samples. The distribution of LBGs follows the bulk of the parent SFGs population, although we see that the LBG sample starts to miss galaxies as they become redder. The OFG sample is complementary to the LBG sample by selecting bright $F444W < 25$\,mag galaxies missed by the LBG criteria. We note that LBGs and OFGs are not mutually exclusive galaxy types; at the edges of the OFGs selection region, seven galaxies are both LBG and OFG. An example OFG is shown in Fig.~\ref{fig:ofgs_example}.

\begin{figure}
\begin{center}
\includegraphics[width=\columnwidth]{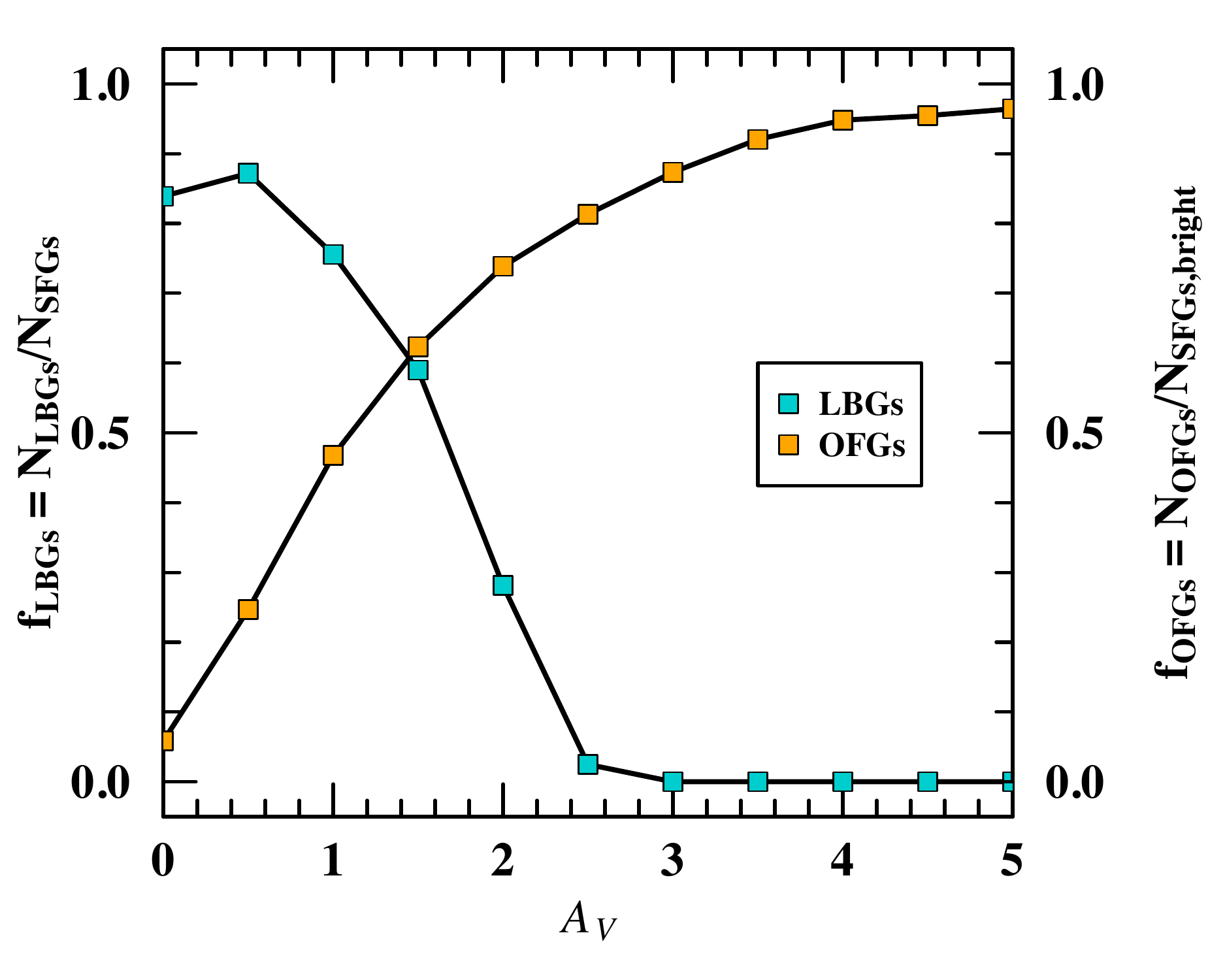}
\caption{Fraction of LBGs (left axis) and fraction of OFGs (right axis) as a function of increasing $A_V$ (see main text, for details on the calculation), where $\rm{N_{SFG}}$ is the number of galaxies in the parent SFG sample and $\rm{N_{SFGs,bright}}$ is the number of galaxies in the parent SFG sample with $F444W < 25$\,mag.}
\label{fig:frac}
\end{center}
\end{figure}

In light of the small percentage of QG contaminants in the OFG sample, we find that most of the $z > 3$ massive galaxies missed by the LBG criteria in the pre-\textit{JWST} era were missed mainly because of high dust attenuation rather than because of old stellar populations. In order to illustrate the effect of dust attenuation in the selection criteria for LBGs and OFGs, we performed the following experiment: First, we took the intrinsic (before applying the dust attenuation law) best-fit SEDs of the parent SFG sample from \texttt{FAST++}. We used the SEDs of the parent SFG sample to build a representative set of SEDs, as opposed to taking all the SED templates of \texttt{FAST++}. We then progressively applied increasing $A_V$ values following the \citet{calzetti00} dust attenuation law. In every $A_V$ step, we obtained the magnitudes in the relevant filters for the selection criteria for LBGs and OFGs by convolving their transmission curves with the best-fit SEDs. In each $A_V$ step, we calculated the fraction of galaxies that meet the selection criteria for LBGs and OFGs, that is, $\rm{f_{LBGs}} = \rm{N_{LBGs}}/\rm{N_{SFGs}}$, where $\rm{N_{SFG}}$ is the number of galaxies in the parent SFG sample; and $\rm{f_{OFGs}} = \rm{N_{OFGs}}/\rm{N_{SFGs,bright}}$, where $\rm{N_{SFG,bright}}$ is the number of galaxies in the parent SFG sample with $F444W < 25$\,mag. Figure~\ref{fig:frac} presents these fractions as a function of increasing $A_V$. This illustrates how the Lyman break technique naturally misses galaxies as a function of increasing $A_V$ ($\sim 50$\% for $A_V = 2$\,mag and $\sim 0$\% for $A_V > 3$\,mag). On the contrary, the higher the $A_V,$ the higher the fraction of OFGs ($\sim 50$\% for $A_V = 1$\,mag and $\sim 100$\% for $A_V > 5$\,mag).

\section{Stellar-based properties and morphologies} \label{sec:prop}

\begin{figure*}
\begin{center}
\includegraphics[width=0.33\textwidth]{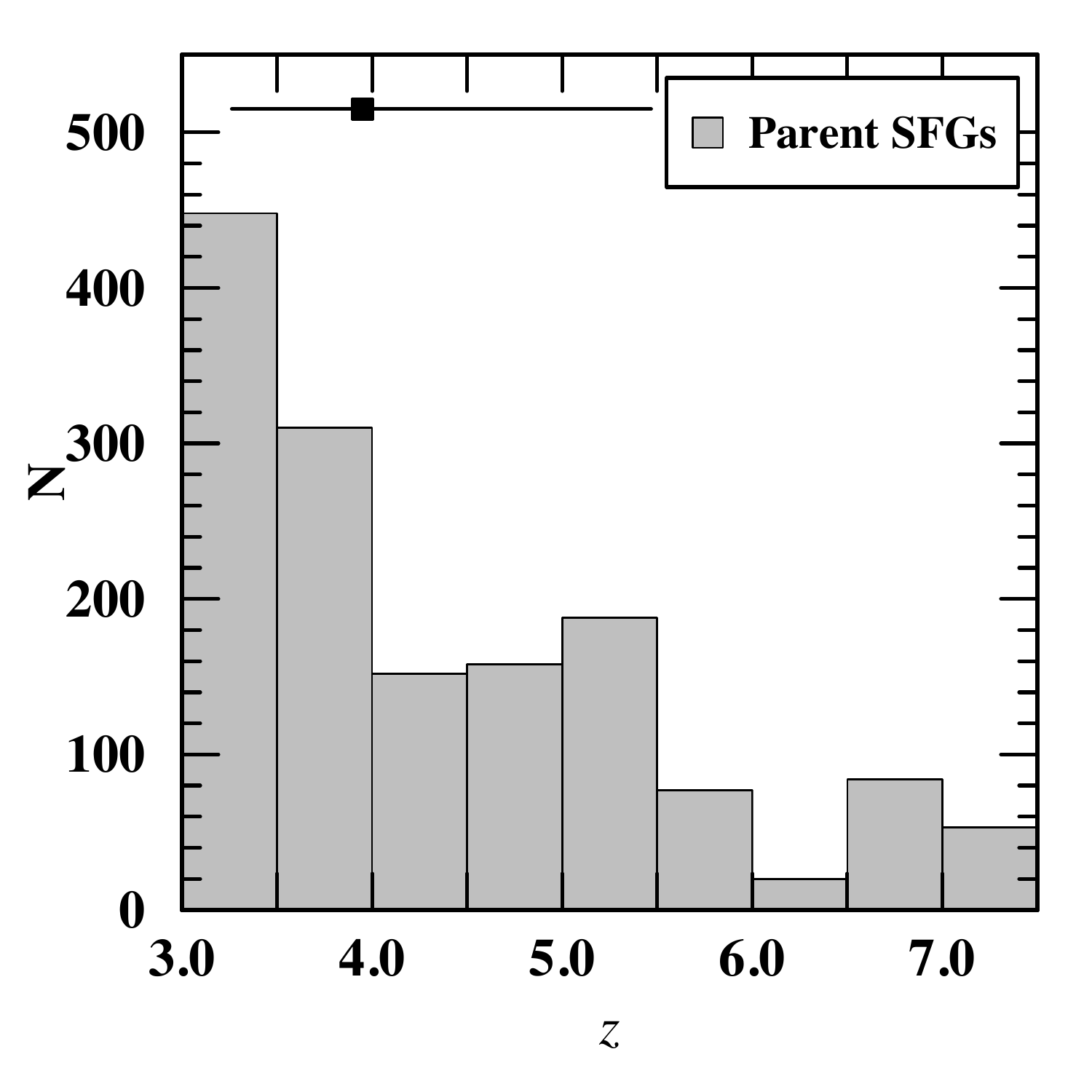}
\includegraphics[width=0.33\textwidth]{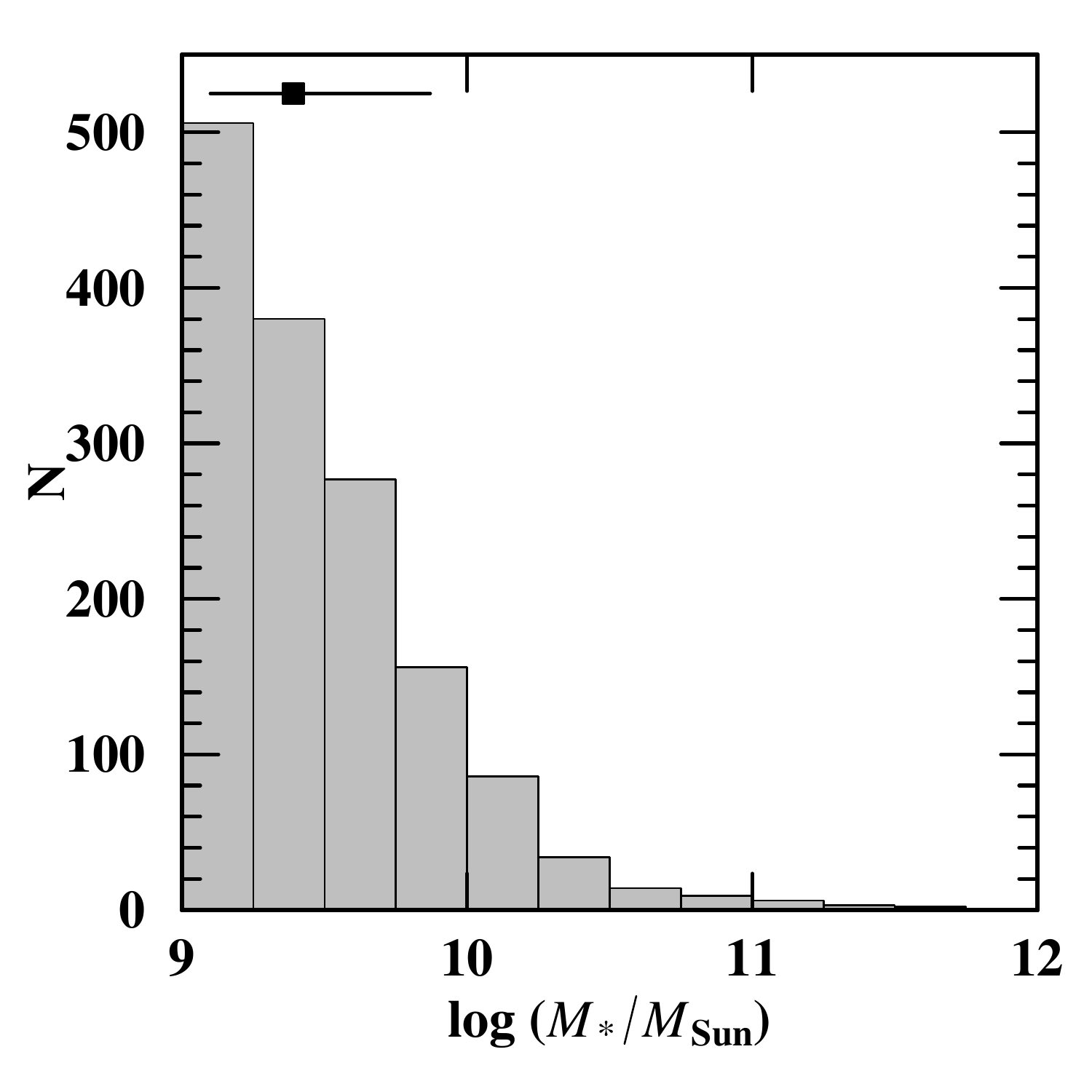}
\includegraphics[width=0.33\textwidth]{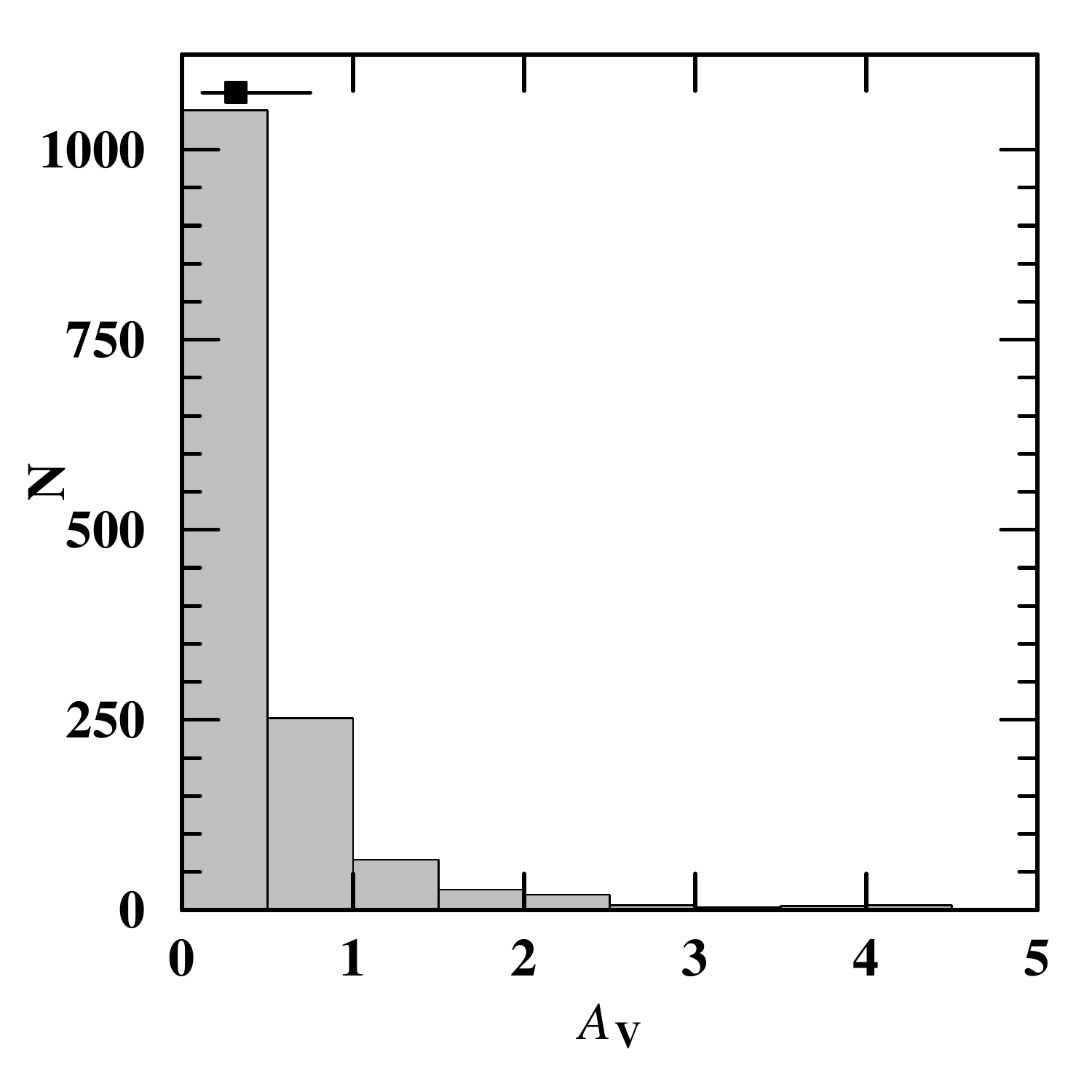}
\includegraphics[width=0.33\textwidth]{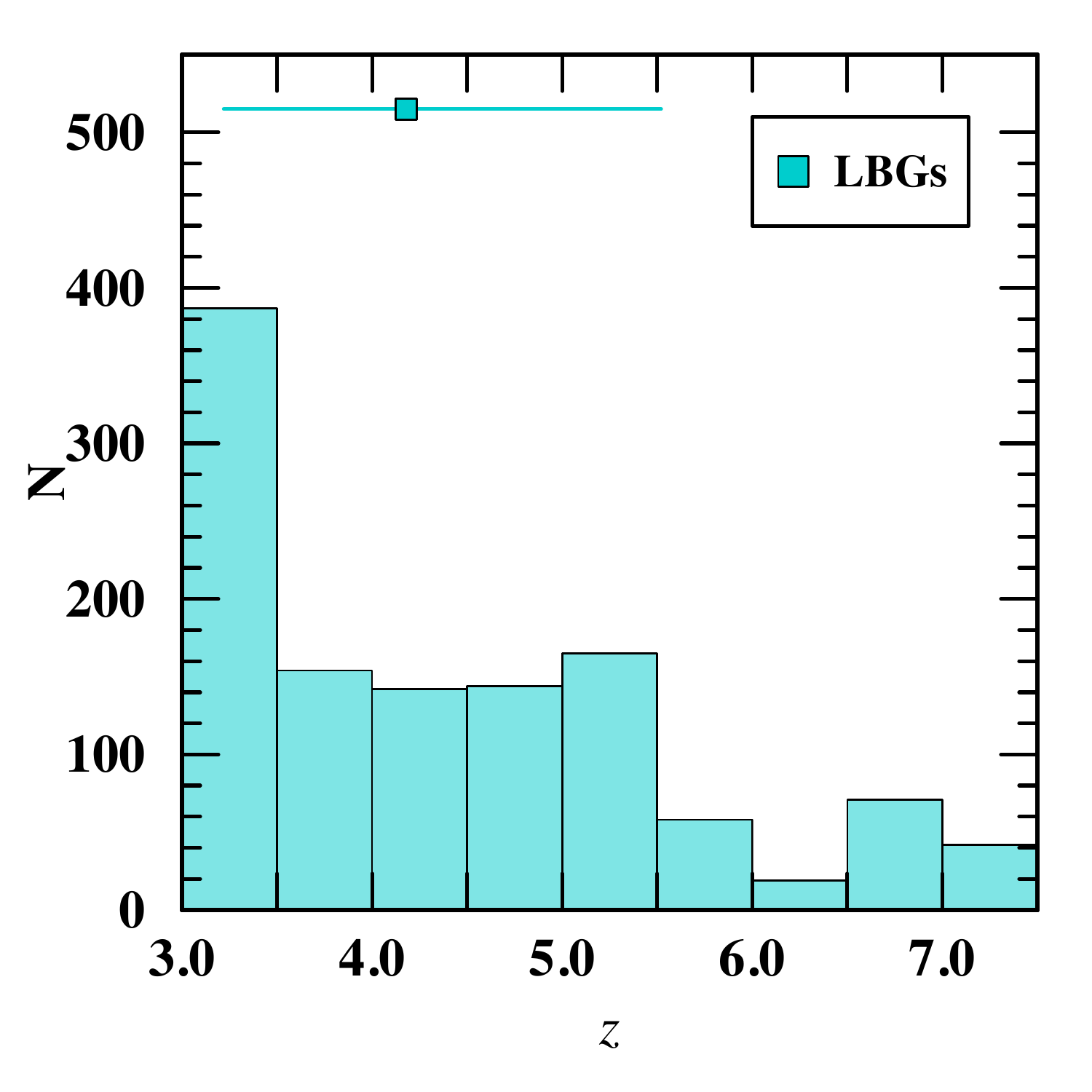}
\includegraphics[width=0.33\textwidth]{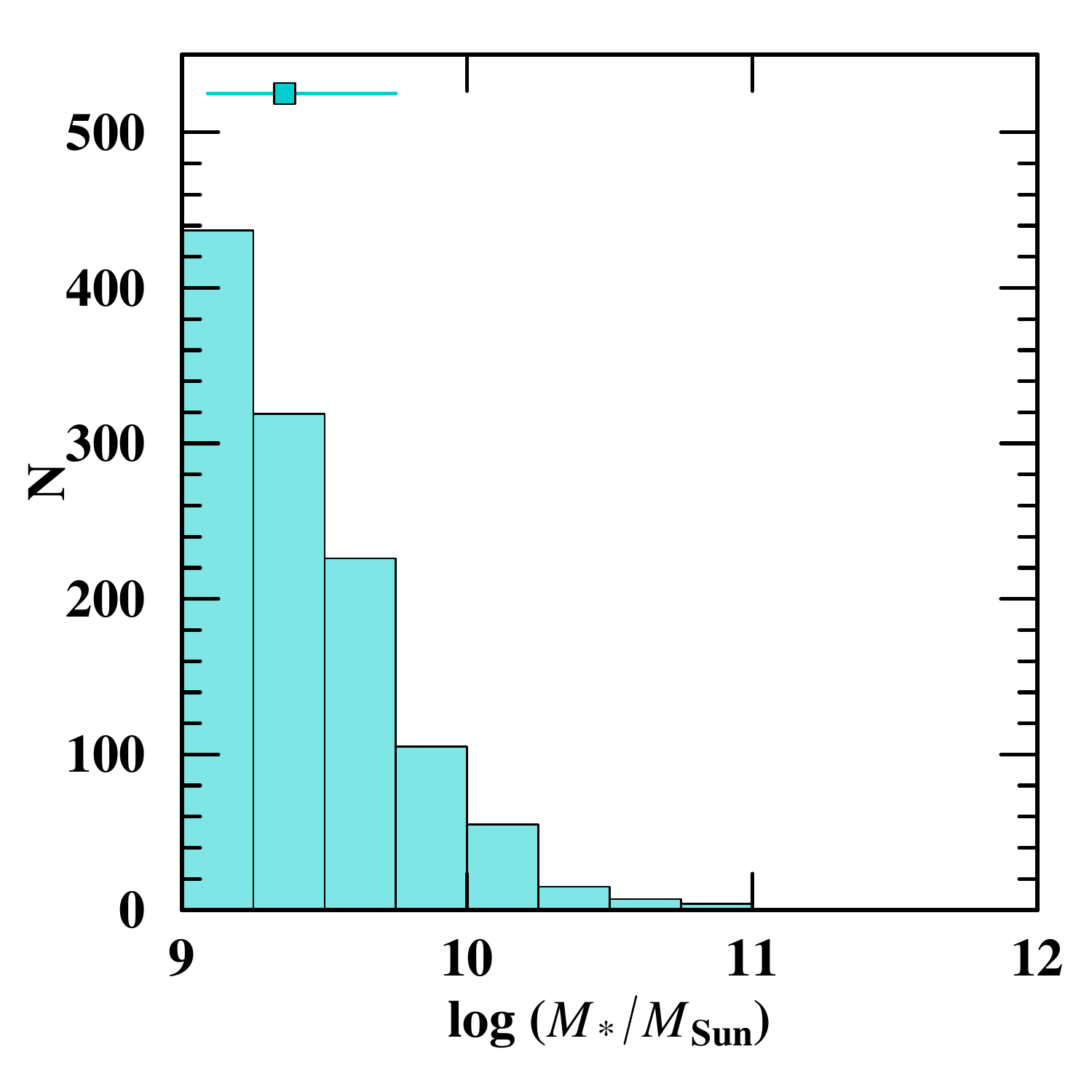}
\includegraphics[width=0.33\textwidth]{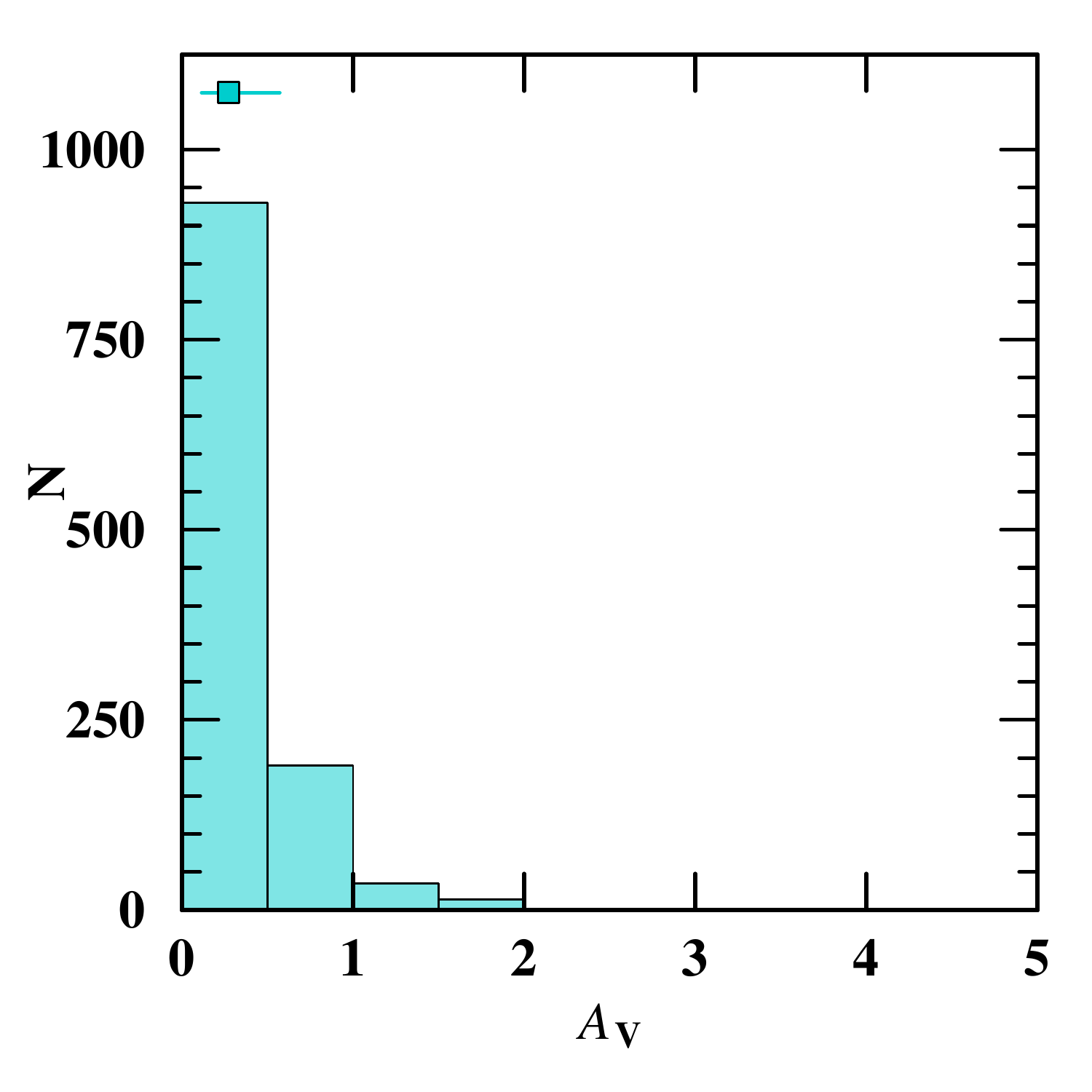}
\includegraphics[width=0.33\textwidth]{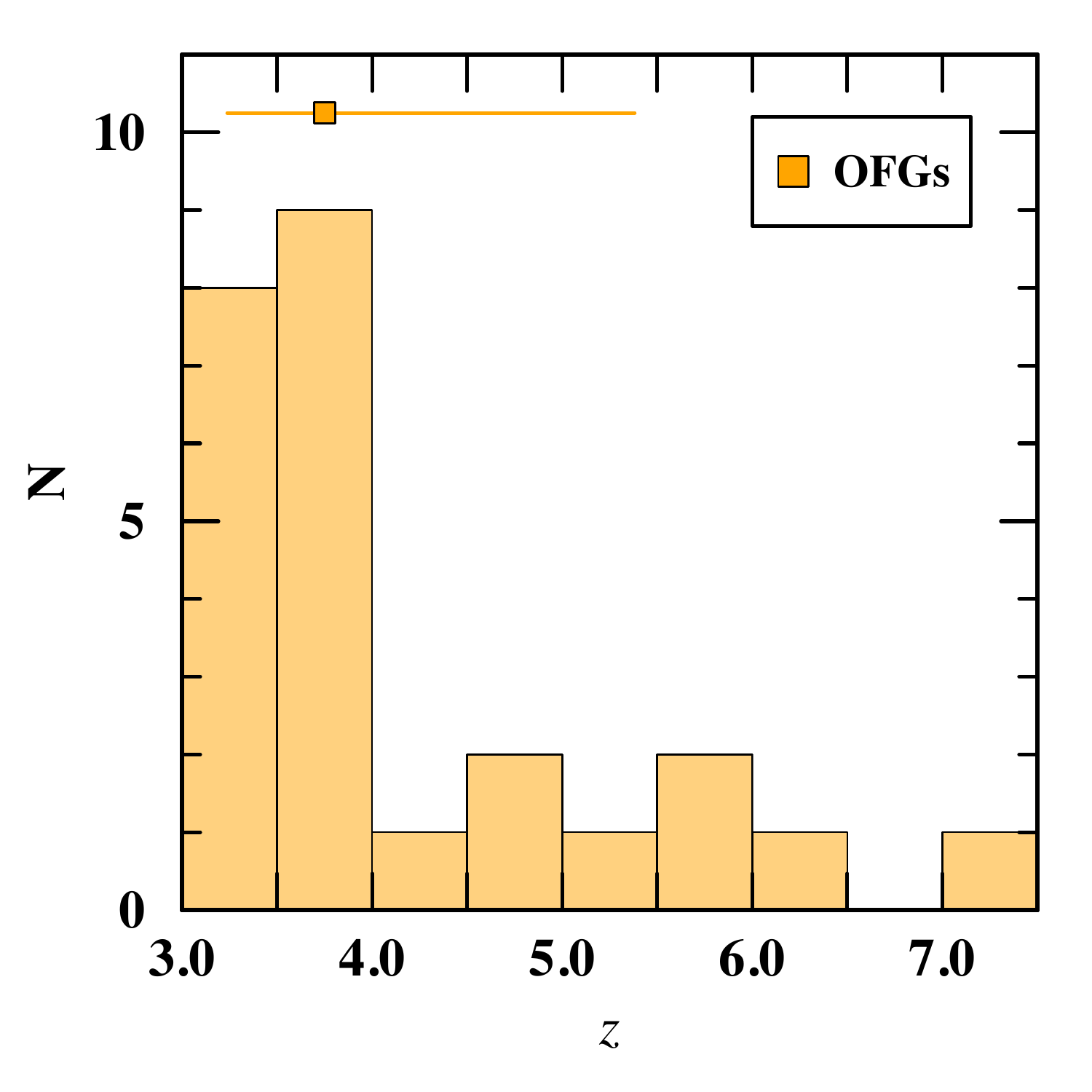}
\includegraphics[width=0.33\textwidth]{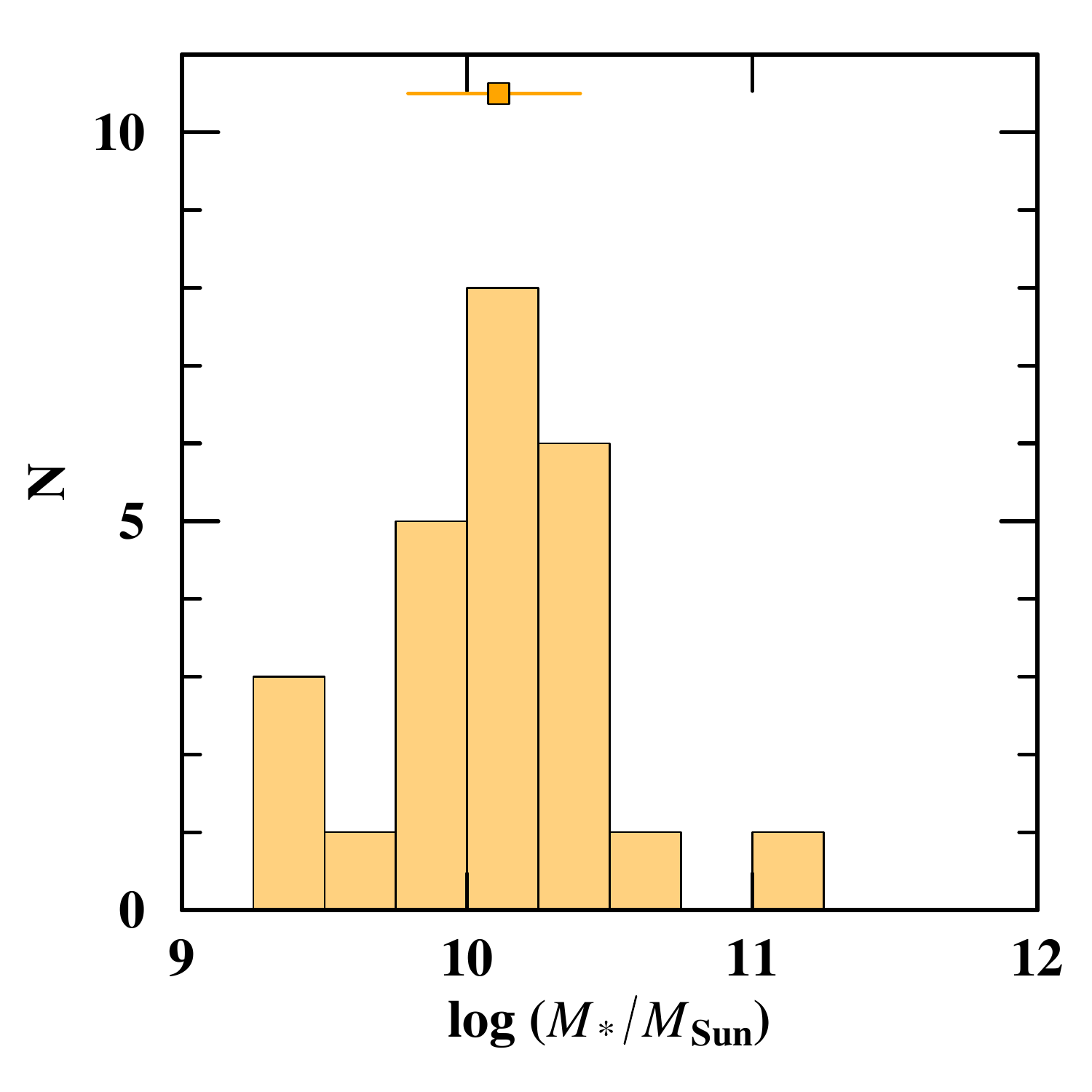}
\includegraphics[width=0.33\textwidth]{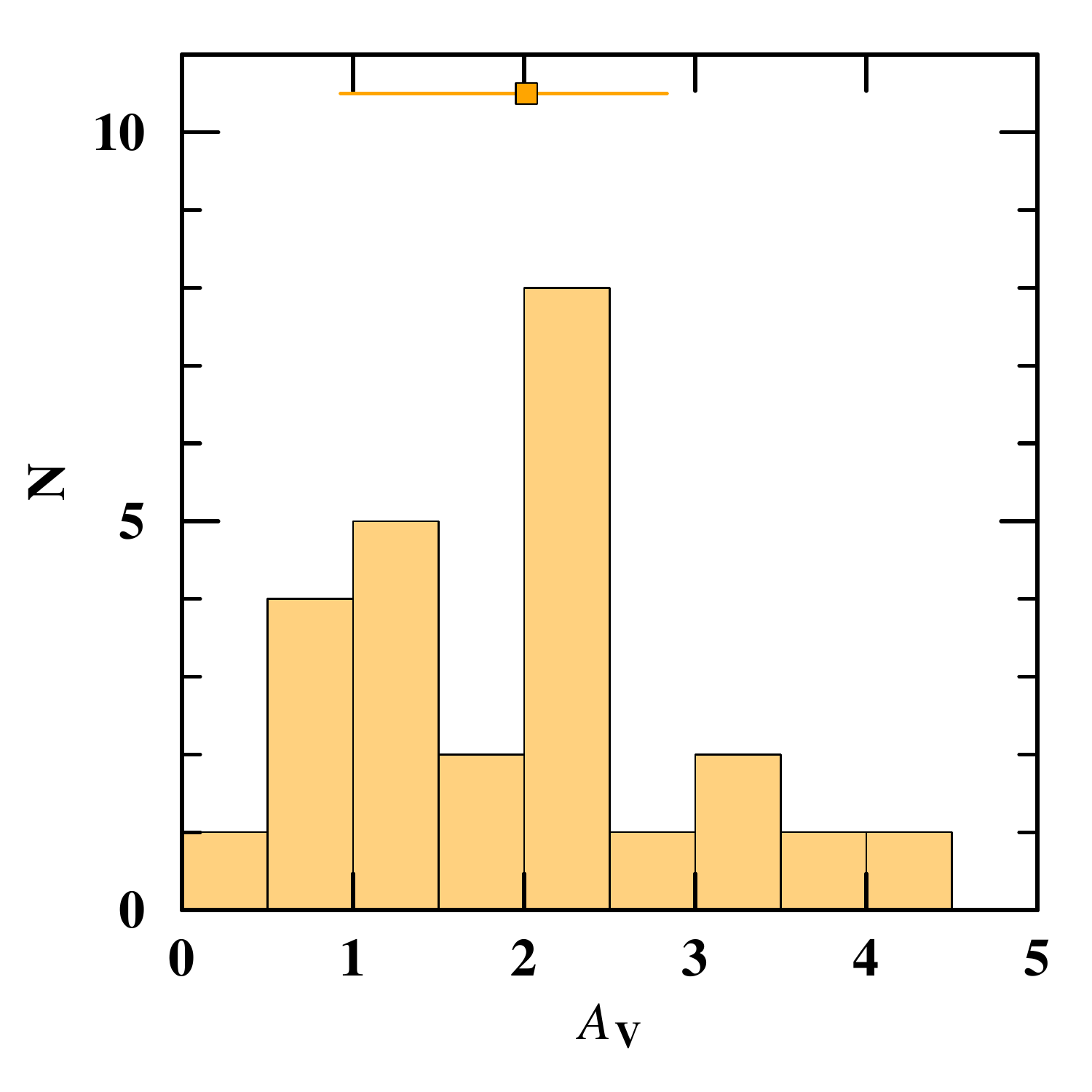}
\caption{Redshift (left column), stellar mass (middle column), and dust attenuation (right column) histograms of the parent SFG (top row, gray), LBG (middle row, cyan), and OFG (bottom row, orange) samples. We show the median values (filled squares) along with the 16\% and 84\% percentiles (segments) of the distributions in the insets.}
\label{fig:hist_z_mstar_av}
\end{center}
\end{figure*}

In this section, we characterize the stellar-based properties, including redshift, stellar mass, dust attenuation, SFR, and morphological measurements, as derived in Sect.~\ref{sec:catalog}. We compare the samples defined as in the previous section with regard to these stellar properties. In Fig.~\ref{fig:hist_z_mstar_av}, we present the redshift, stellar mass, and dust attenuation distributions of the parent SFG, LBG, and OFG samples to illustrate the parameter space covered by these galaxies. All samples exhibit a similar redshift distribution. In terms of stellar mass and dust attenuation distributions, the main difference is in the OFG sample. These galaxies are more massive and dust attenuated than LBGs and the parent sample of SFGs. This reflects the OFG criteria, which are designed to select intermediate to massive galaxies with moderate to high levels of dust attenuation.

\begin{figure}
\begin{center}
\includegraphics[width=\columnwidth]{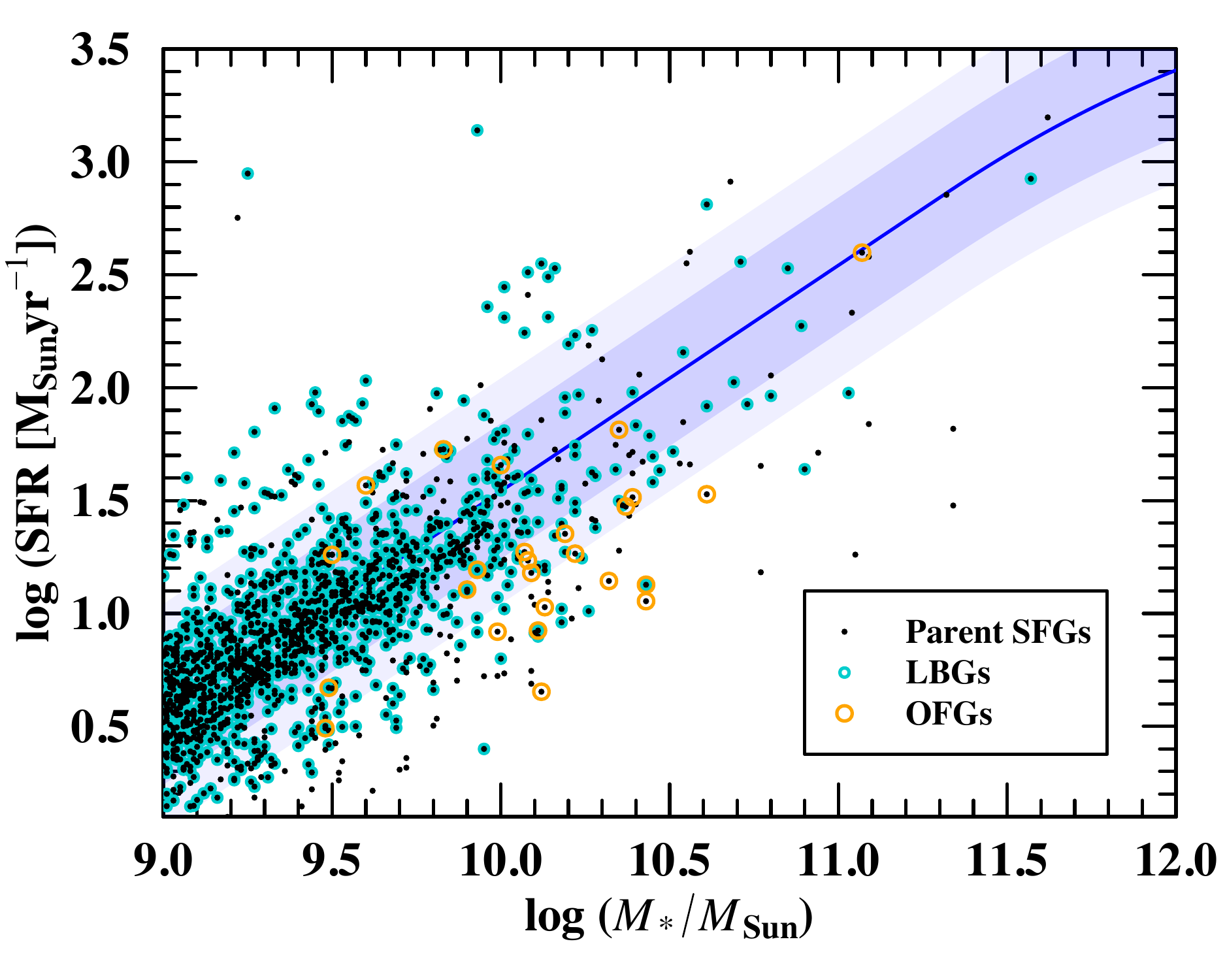}
\caption{SFR--$M_{\rm{*}}$ plane, with the MS from \citet{schreiber15} displayed as a solid blue line. Its 1$\sigma$ scatter associated with $0.5 < \Delta \rm{MS} < 2$ ($\sim 0.3$\,dex) is represented as a shaded blue area, with a more extended typical scatter of $0.33 < \Delta \rm{MS} < 3$ ($\sim 0.5$\,dex) in lighter blue. We note that the values are scaled to a common redshift ($z_{\rm{med}} = 3.95$) as explained in the main text.}
\label{fig:ms}
\end{center}
\end{figure}

In Fig.~\ref{fig:ms}, we place our different samples in the stellar mass versus SFR plane, along with the MS of SFGs as defined by \citet{schreiber15}, assuming that this MS definition can be extrapolated to $z > 4$. The SFR values are scaled to a common redshift, corresponding to the median value of the parent SFGs sample ($z_{\rm{med}} = 3.95$). The scaled SFR values do not represent the true SFR values, but the overall distribution of the different samples in the SFR--$M_{\rm{*}}$ plane, maintaining the $\Delta \rm{MS} = \rm{SFR/SFR_{MS}}$ each galaxy would have if plotted against the MS associated with its redshift. The parent SFG sample has a median $\Delta \rm{MS} = 0.91_{-0.40}^{+0.74}$ (where the uncertainties are the 16\% and 84\% percentiles of the distributions), tracing the MS of SFGs. Similarly, LBGs have $\Delta \rm{MS} = 0.95_{-0.37}^{+0.67}$ and therefore trace the bulk of the MS of SFGs. In the case of OFGs, $\Delta \rm{MS} = 0.38_{-0.19}^{+0.66}$ locates them slightly below the MS of SFGs. Previous studies revealed that the SFR of the OFGs (or comparable selections)  is consistent with the typical MS SFGs as constrained from stacked mid-IR-to-mm photometry or deep mm individual detections from ALMA \citep[e.g.,][]{wang19,gomezguijarro22b,xiao23}. SFR could be underestimated in OFGs in the absence of counterparts in the mid-IR-to-mm that typically offer the best SFR constraints. In this work, none of the OFGs have counterparts in the mid-IR-to-mm bands in the super-deblended catalog in the EGS field, and so SFRs would be underestimated, likely because of an underestimated dust attenuation, which already reaches saturation in the SED fitting of such highly dust-attenuated systems. The extreme cases are the OFGs that are still undetected in their rest-frame UV by \textit{JWST} (9/25), for which dust attenuation constraints are heavily dependent on extrapolations of the best-fit SED. These galaxies do not show differences in terms of redshift when compared to the OFG sample ($\sim 5$\% difference in the medians) and exhibit 1.7 times higher stellar masses and 1.2 times  higher dust attenuation. \citet{perezgonzalez23} reported a similar SFR behaviour compared to the MS for \textit{HST}-dark galaxies, which is mitigated when performing 2D SED-fitting.

\begin{figure*}
\begin{center}
\includegraphics[width=0.33\textwidth]{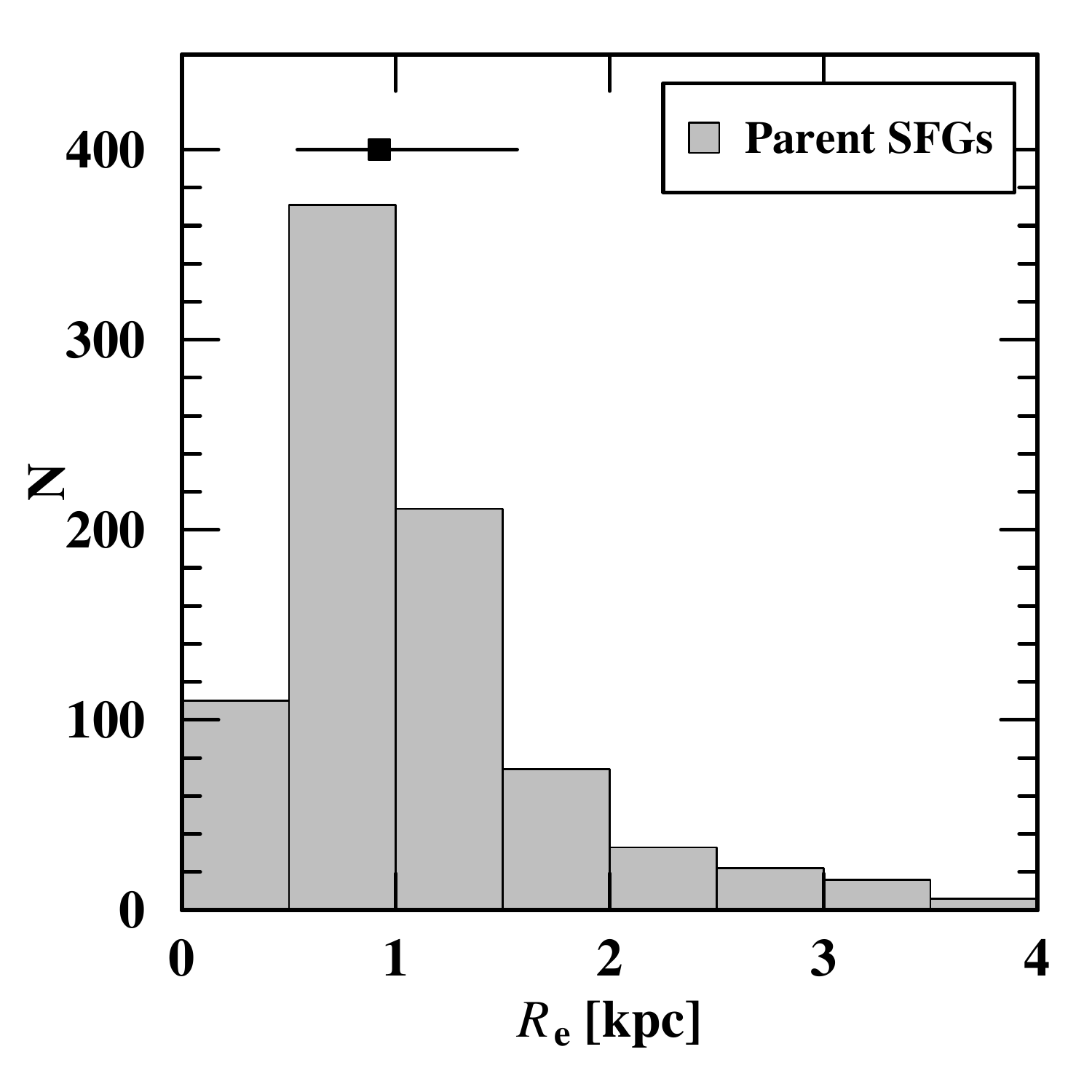}
\includegraphics[width=0.33\textwidth]{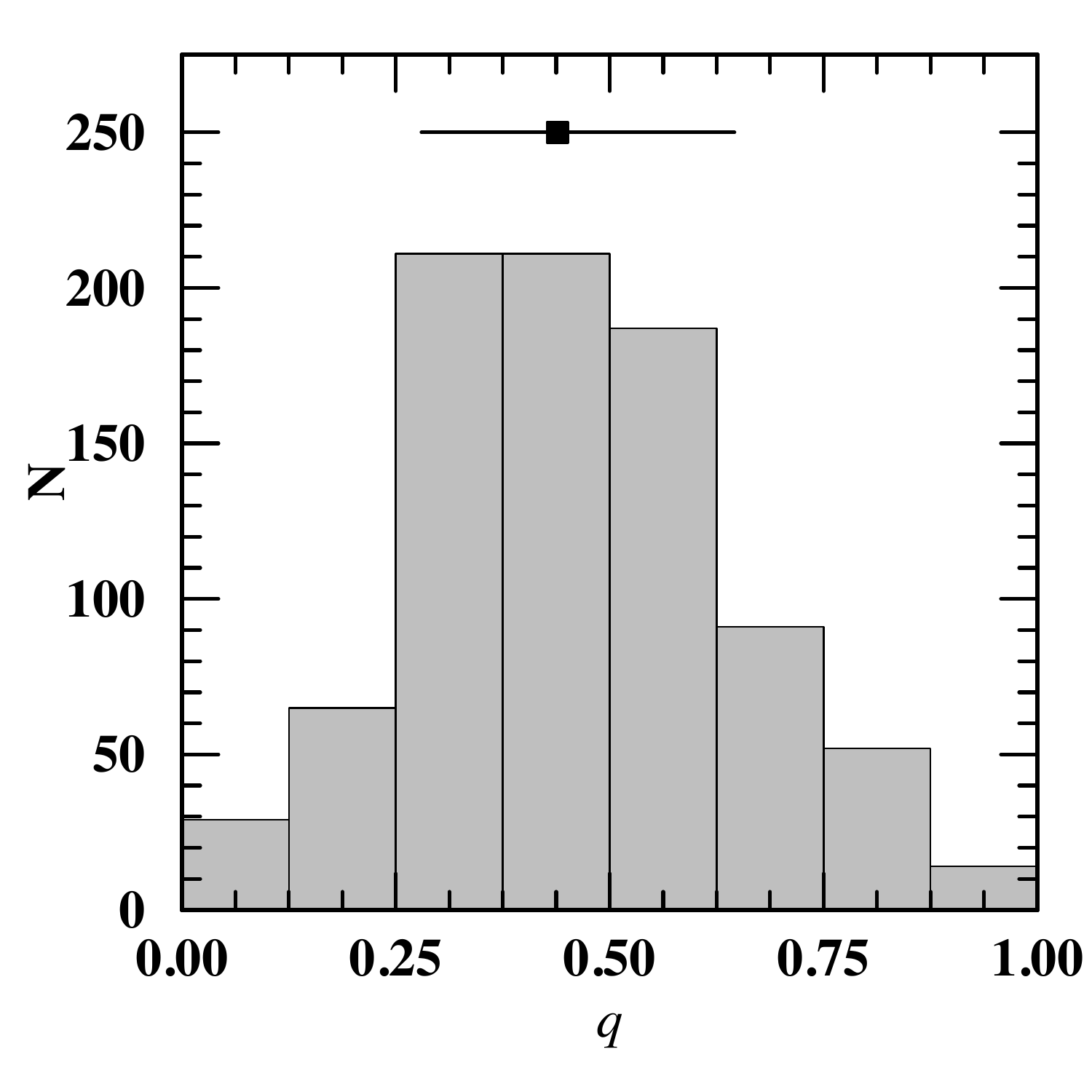}
\includegraphics[width=0.33\textwidth]{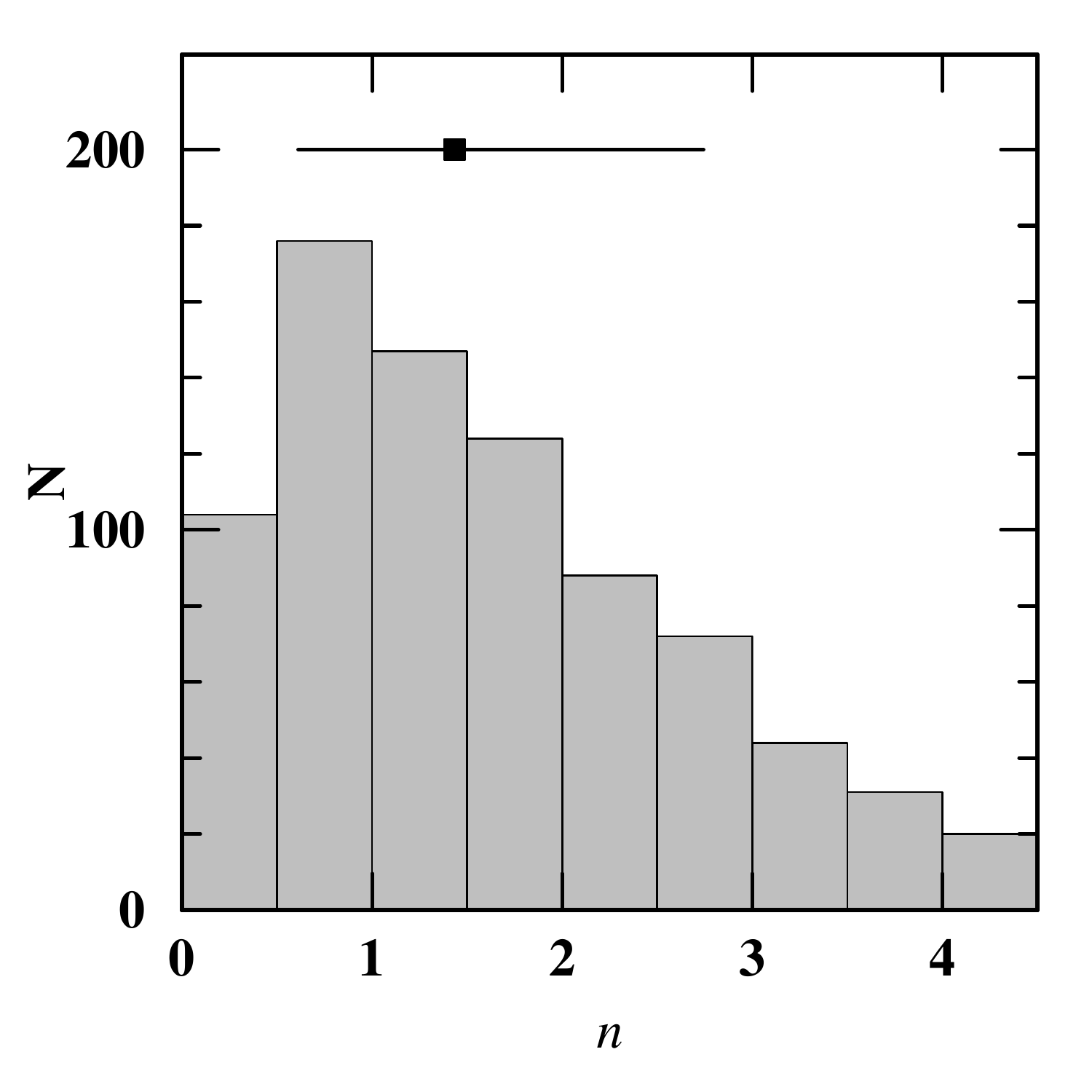}
\includegraphics[width=0.33\textwidth]{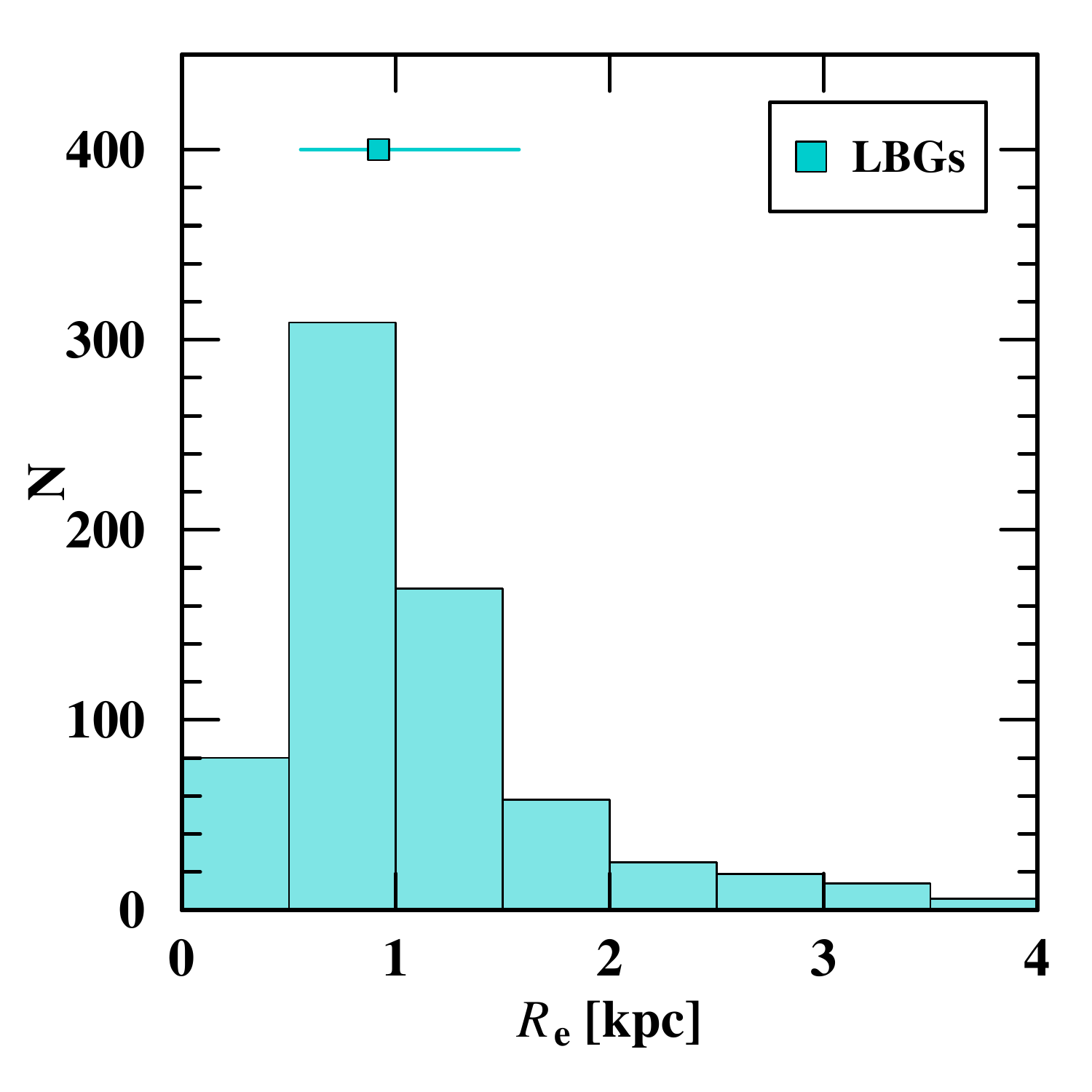}
\includegraphics[width=0.33\textwidth]{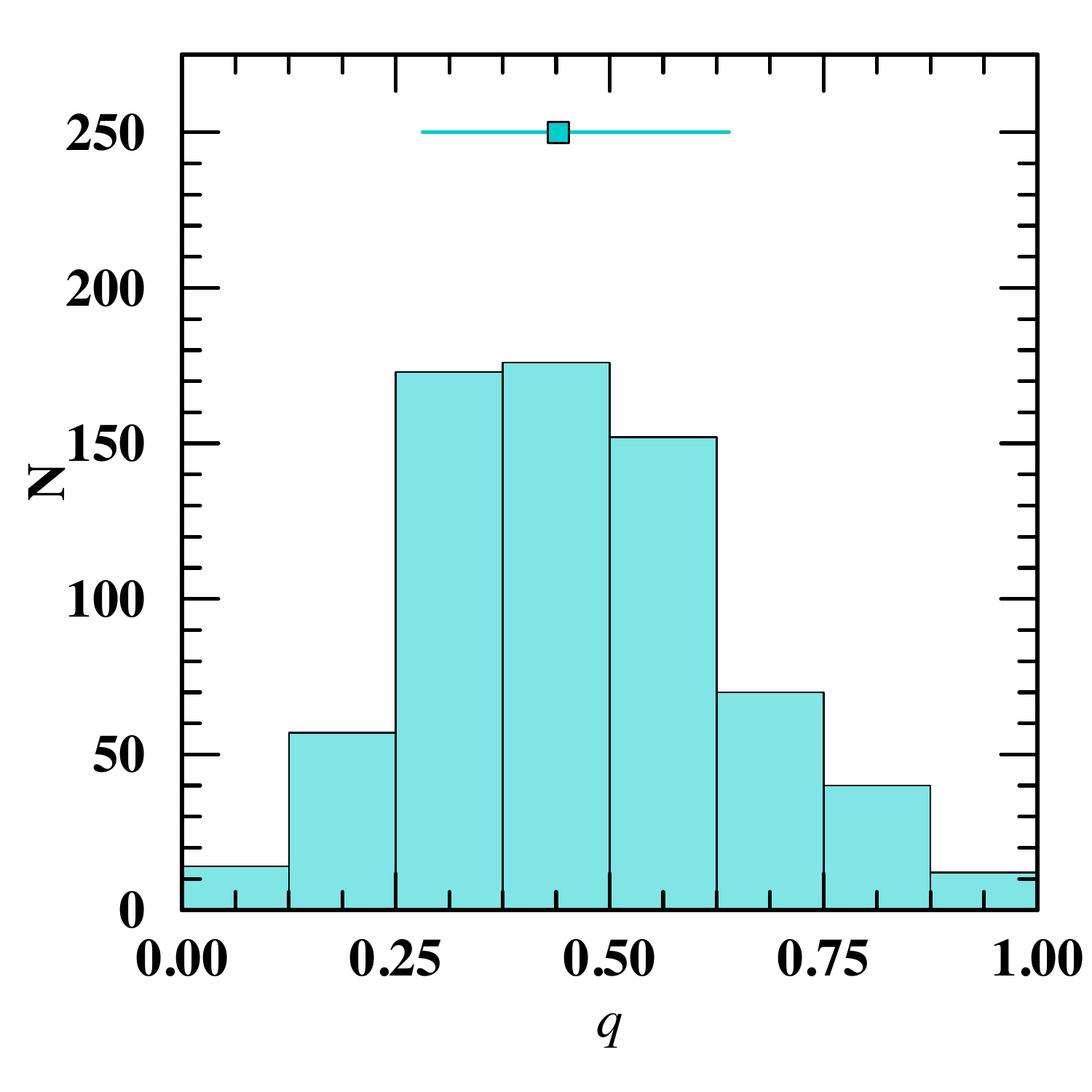}
\includegraphics[width=0.33\textwidth]{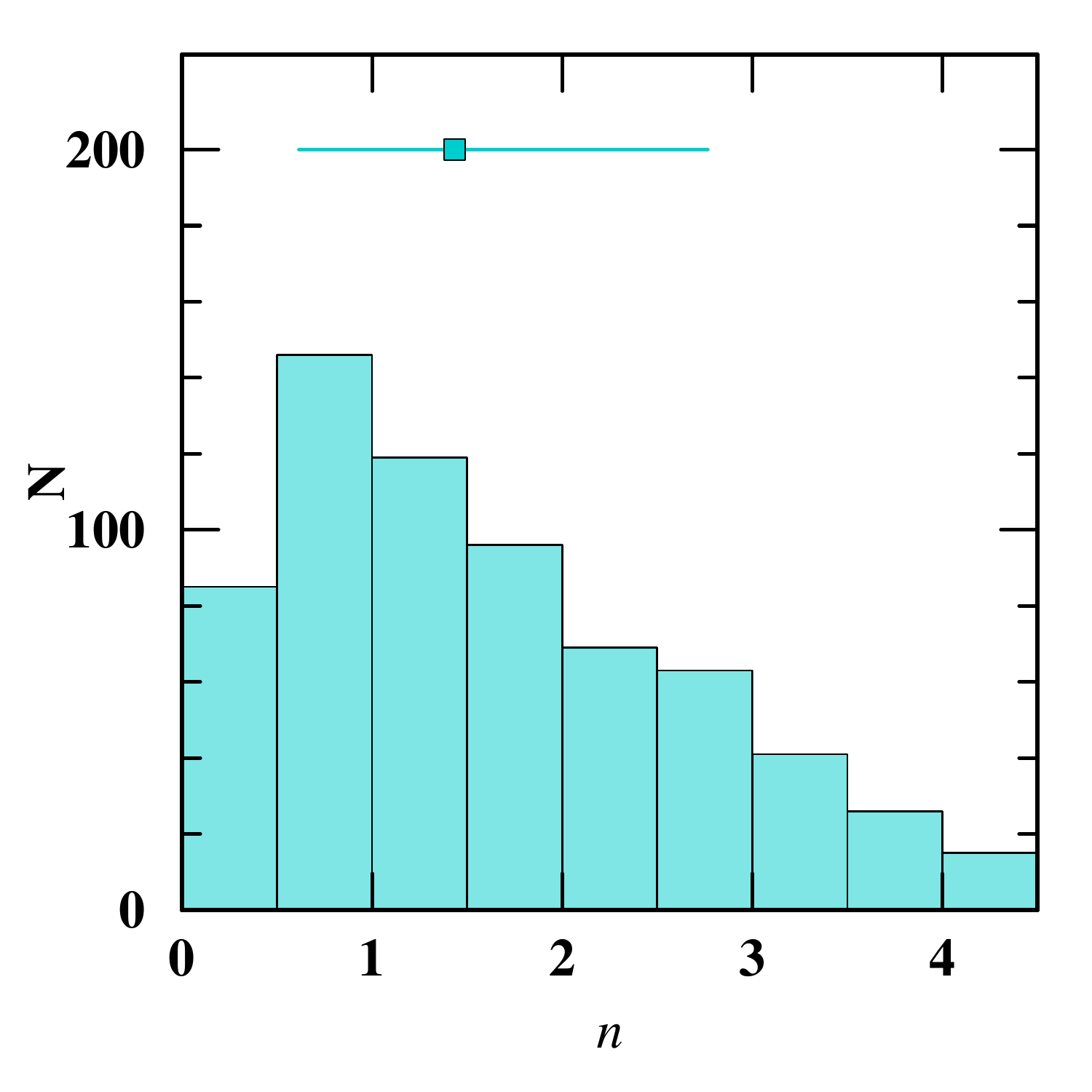}
\includegraphics[width=0.33\textwidth]{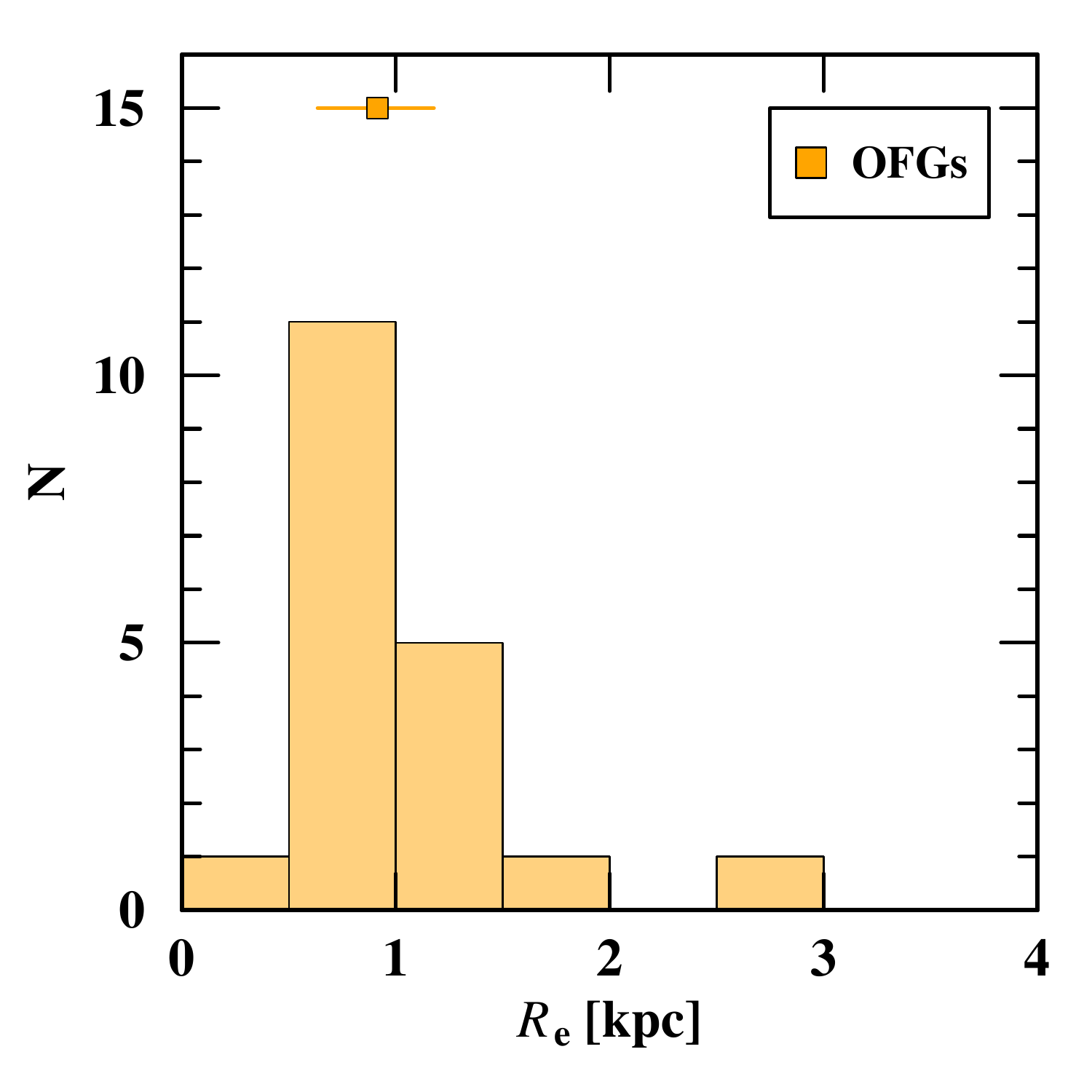}
\includegraphics[width=0.33\textwidth]{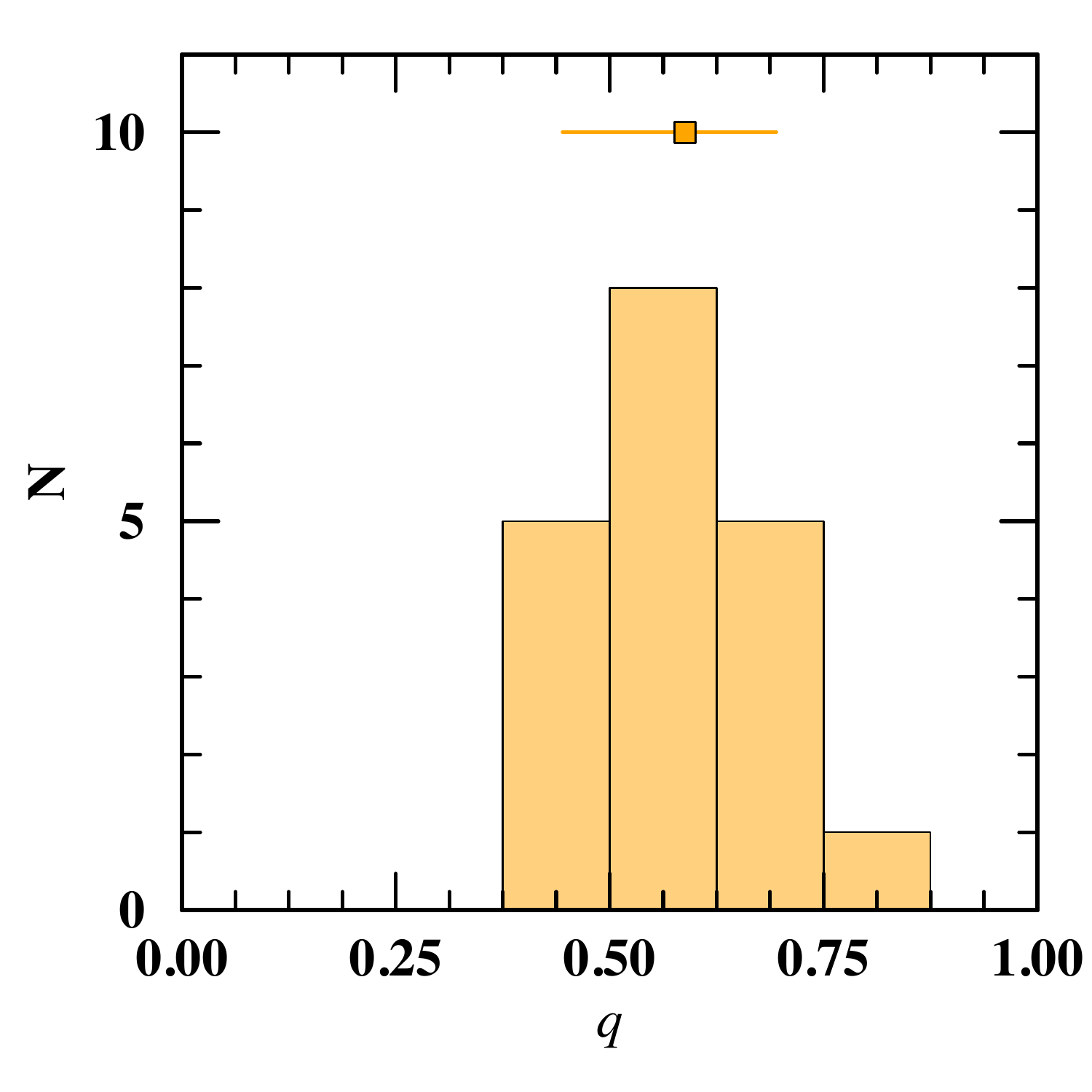}
\includegraphics[width=0.33\textwidth]{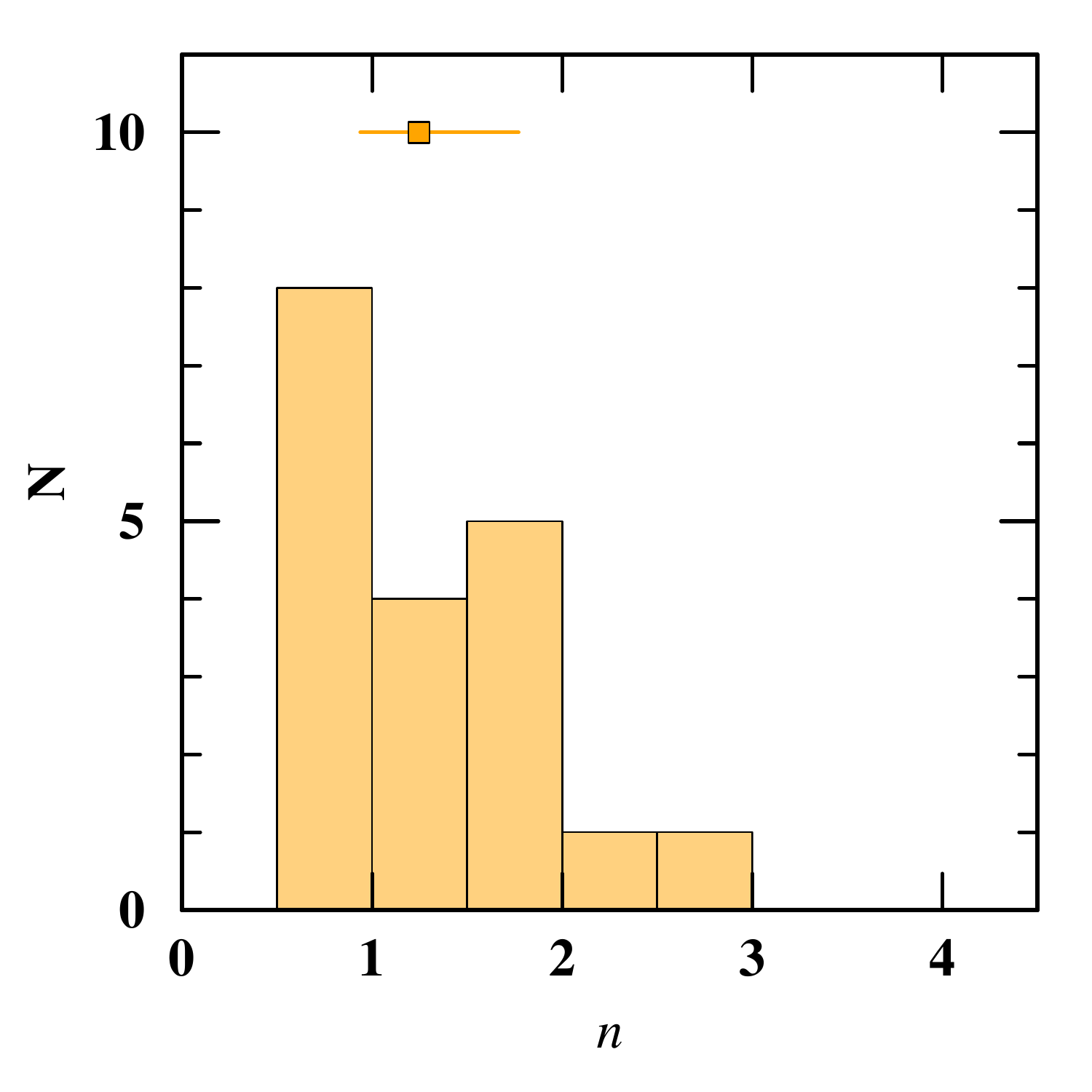}
\caption{(Circularized) effective radius (left column), axis ratio (middle column), and S\'ersic index (right column) histograms of the parent SFG (top row, grey), LBG (middle row, cyan), and OFG (bottom row, orange) samples. We show the median values (filled squares) along with the 16\% and 84\% percentiles (segments) of the distributions  in insets.}
\label{fig:hist_morph}
\end{center}
\end{figure*}

In terms of morphological parameters, in Fig.~\ref{fig:hist_morph}, we present the (circularized) effective radius ($R_{\rm{e}}$), axis ratio ($q = b/a$), and S\'ersic index ($n$) distributions of the parent SFG, LBG, and OFG samples (flagged as good fits). We recall that these were measured in the \textit{JWST}/NIRCam $F444W$-band and that we restricted our analysis to $flag \leq 1$ and $flag\_sersic = 0$ when considering morphological measurements. Among the OFGs sample, 6/25 galaxies were flagged as poor fits, in all cases being found to be consistent with point sources. Therefore, while it was not possible in these cases to measure their morphologies, a limit on their effective radii is $\rm{PSF_{FHWM}} / 2 < 0\farcs08$. At first order, the distributions are rather similar in the different samples. The main difference is in the lower scatter around small effective radii and S\'ersic index $n \sim 1$ in the case of the OFGs sample compared to LBGs and the parent SFGs sample. In addition, the distribution of axis ratios appears skewed toward larger values in the case of OFGs. While not displayed here, the upper limits on the sizes for the 6/25 OFGs classified as point sources are also consistent with these sources being clustered around smaller effective radii. Their compact sizes could also explain the skewness toward large axis ratios, as the latter are more difficult to constrain when the angular resolution starts to be comparable with the intrinsic size of the galaxy (marginally resolved sources).

\section{Drivers of dust attenuation} \label{sec:av_drivers}

In this section, we study the relevance of stellar mass and morphology in driving dust attenuation in a general sense and, specifically, in the case of highly dust attenuated galaxies, such as OFGs. We note that drivers in this case refer to parameters that show correlations, and these can therefore be interpreted as predictors of the dust attenuation. For a discussion on the relevant physical processes involved, we refer to Sect.~\ref{sec:discussion}.

\subsection{Stellar mass} \label{subsec:dust_mstar}

\begin{figure*}
\begin{center}
\includegraphics[width=0.33\textwidth]{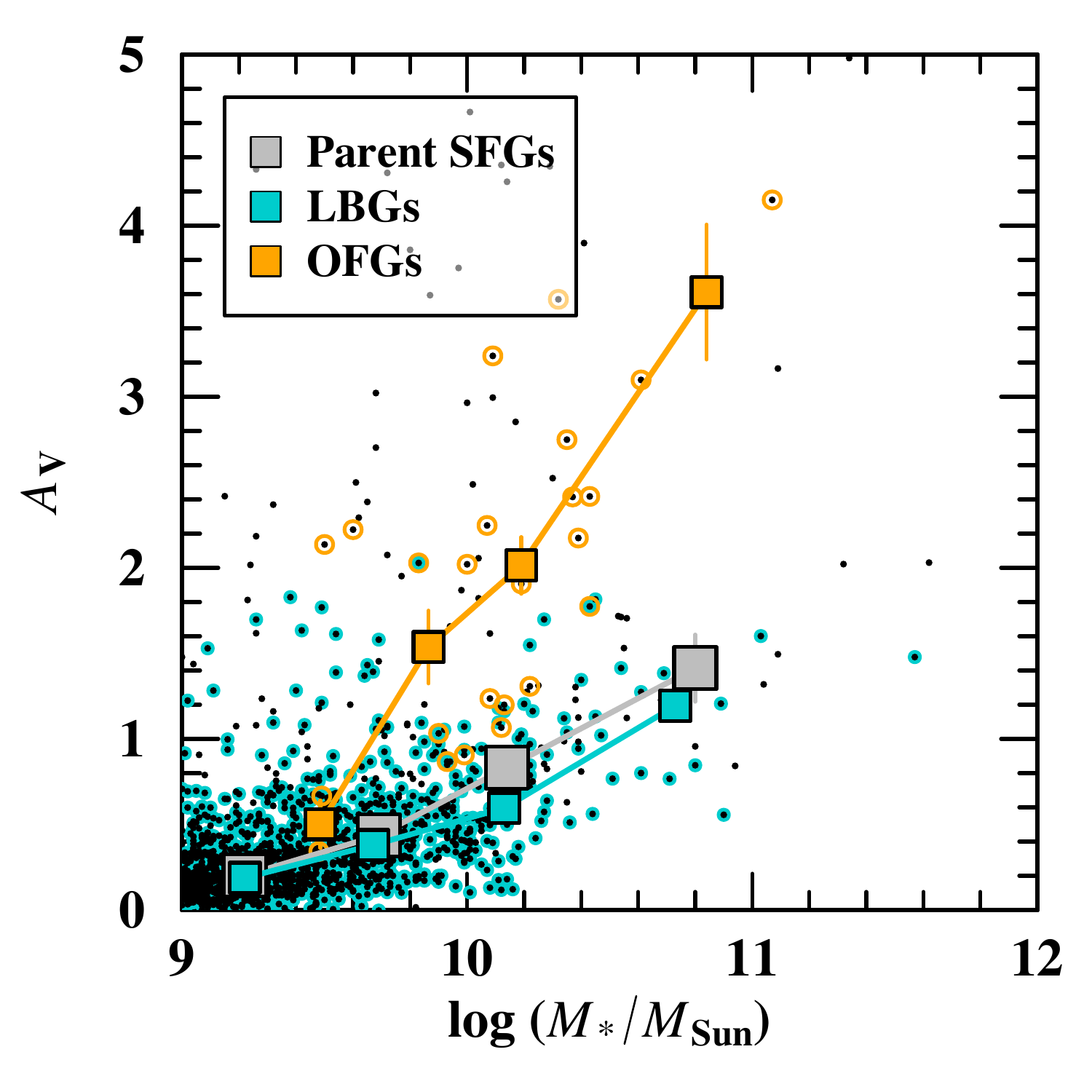}
\includegraphics[width=0.33\textwidth]{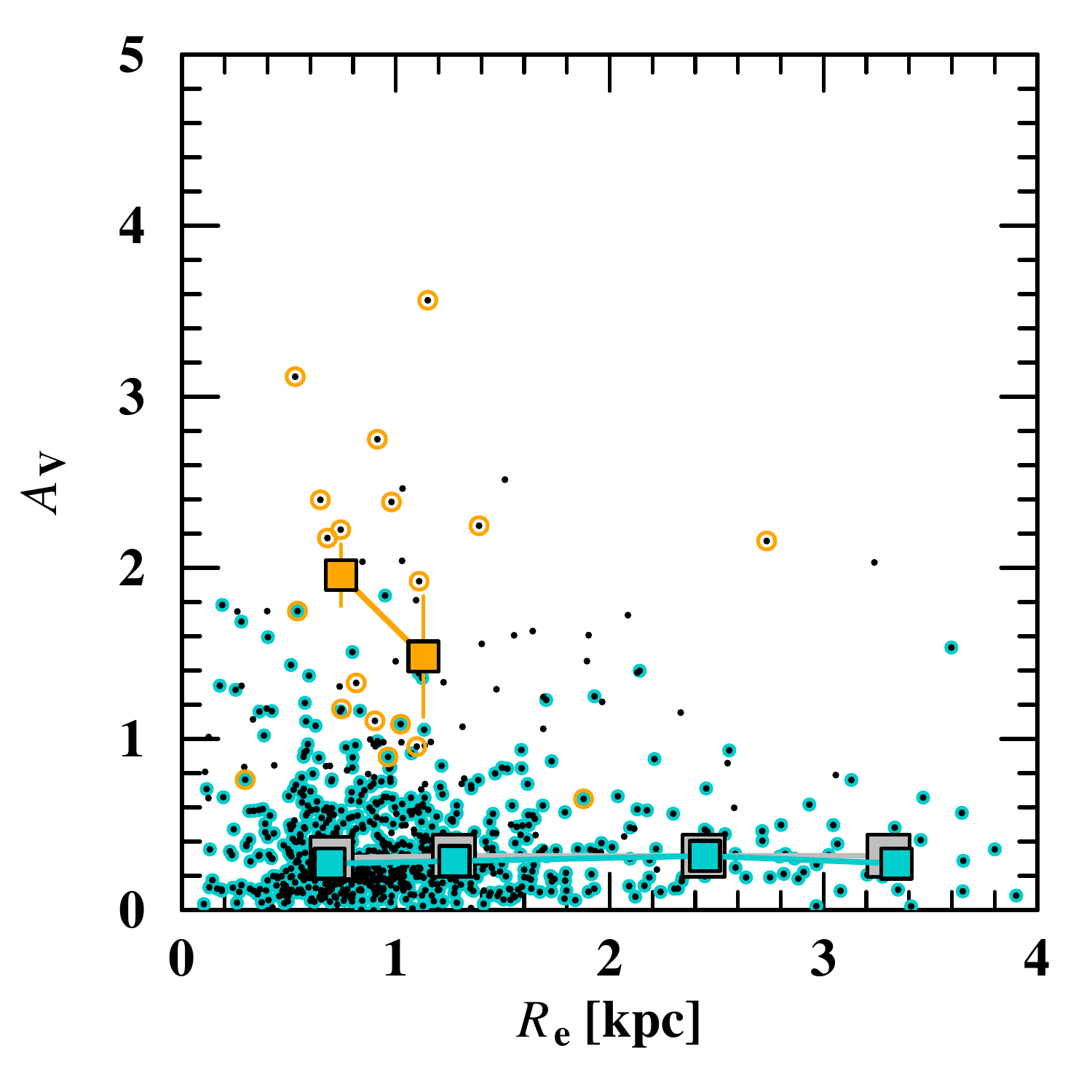}
\includegraphics[width=0.33\textwidth]{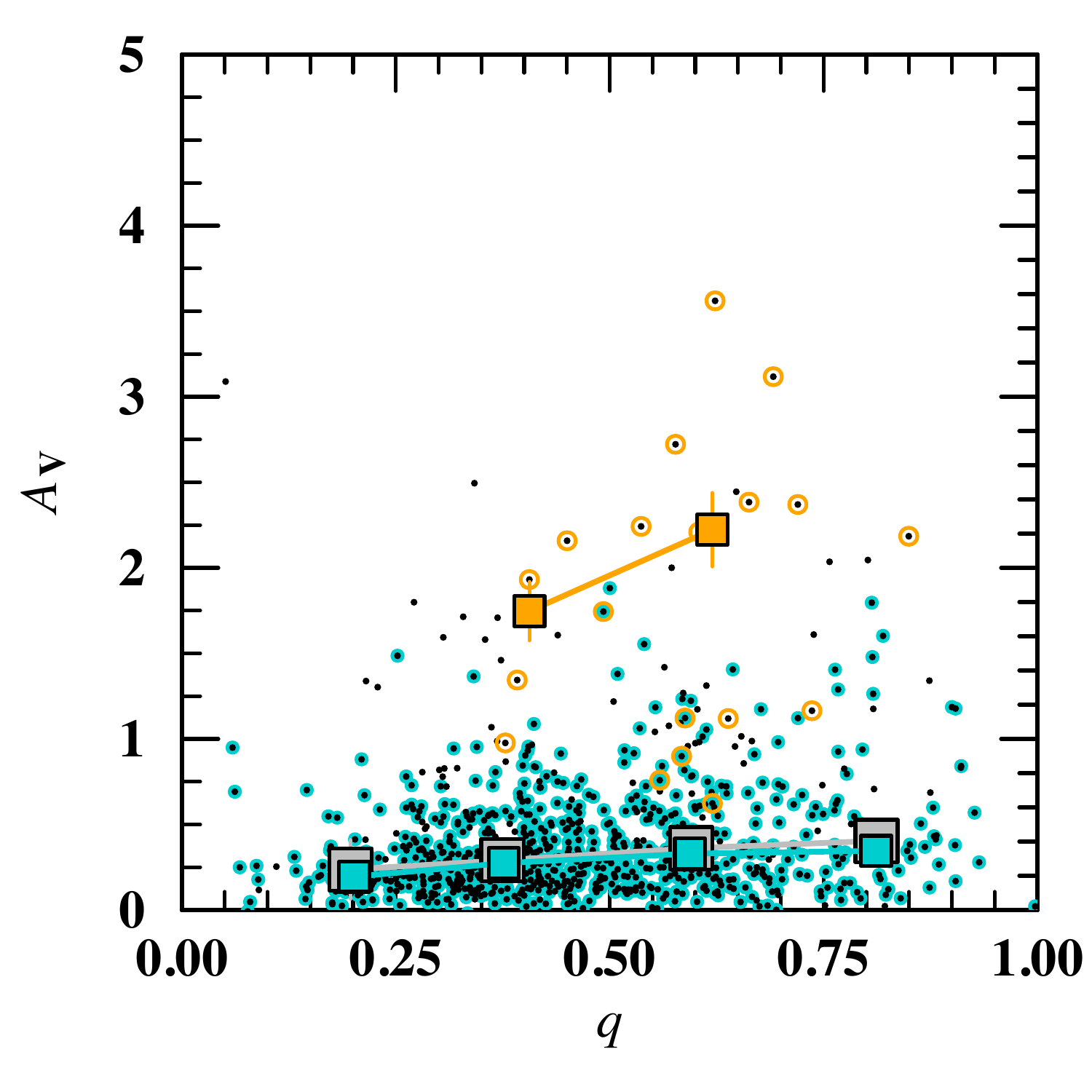}
\caption{Dust attenuation as a function of stellar mass (left panel), (circularized) effective radius (middle panel), and axis ratio (right panel). Sliding medians for the different galaxy types shown in the legend are displayed with colored squares, with the error bars representing the uncertainty of the medians. Bins with less than three galaxies are not displayed (see Figs.~\ref{fig:hist_z_mstar_av} and \ref{fig:hist_morph} for the distributions).}
\label{fig:av_mstar_morph}
\end{center}
\end{figure*}

Figure~\ref{fig:av_mstar_morph} shows dust attenuation as a function of stellar mass. Dust attenuation correlates with stellar mass in SFGs, as reported before in numerous studies \citep[e.g.,][]{garn10,zahid13,heinis14,pannella15,alvarezmarquez16}. The parent sample of SFGs establishes the general trend. LBGs exhibit a shallower trend, while OFGs display a steeper trend populating a much more dust attenuated regime.

The dust attenuation and stellar mass correlation arises from the mass--metallicity relation \citep[e.g.,][]{tremonti04,erb06,sanchez13,genzel15,ma15}. Massive SFGs are capable of more efficiently producing and retaining their metals than low-mass SFGs, which tend to expel them in galactic winds. Therefore, massive SFGs tend to be more dusty than low-mass SFGs.

Stellar mass acts as a primary proxy for dust attenuation in SFGs. However, it remains unclear as to whether or not additional parameters can drive dust attenuation. In the following section, we study the role of morphology.

\subsection{Morphology} \label{subsec:dust_morph}

We studied whether dust attenuation shows signs of correlation with basic morphological structural parameters. In Fig.~\ref{fig:av_mstar_morph}, we show dust attenuation as a function of effective radius and axis ratio. The behavior of $A_V$ for the parent sample of SFGs and LBGs is generally flat across the range of effective radii and axis ratios. The main difference arises in the case of OFG sizes. The effective radii of OFGs are clustered around small sizes (as shown in Fig.~\ref{fig:hist_morph}) that are associated to high dust-attenuation values. In addition, we see an increase in $A_V$ when moving toward smaller effective radii. On the contrary, we do not see an increase in $A_V$ toward lower axis ratios (i.e., edge-on galaxies).

Generally, there is no clear link ---as in the one involving stellar mass--- between the studied basic structural parameters and dust attenuation. However, galaxy size appears to be related in the most dust attenuated galaxy type, the OFG sample. The question that rises is whether or not OFGs are indeed smaller when compared to the typical SFG sizes at similar stellar mass and redshift.

\begin{figure*}
\begin{center}
\includegraphics[width=\columnwidth]{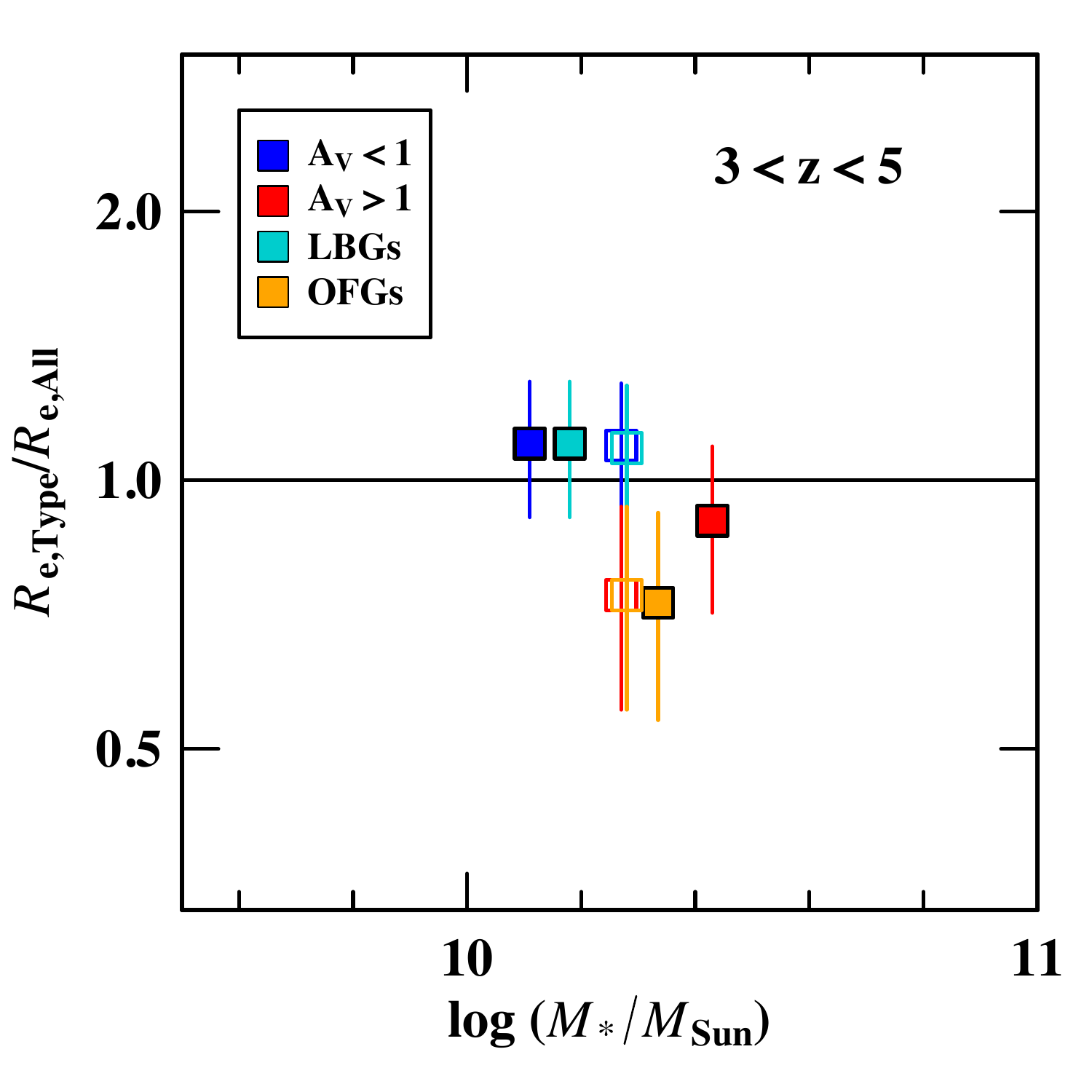}
\includegraphics[width=\columnwidth]{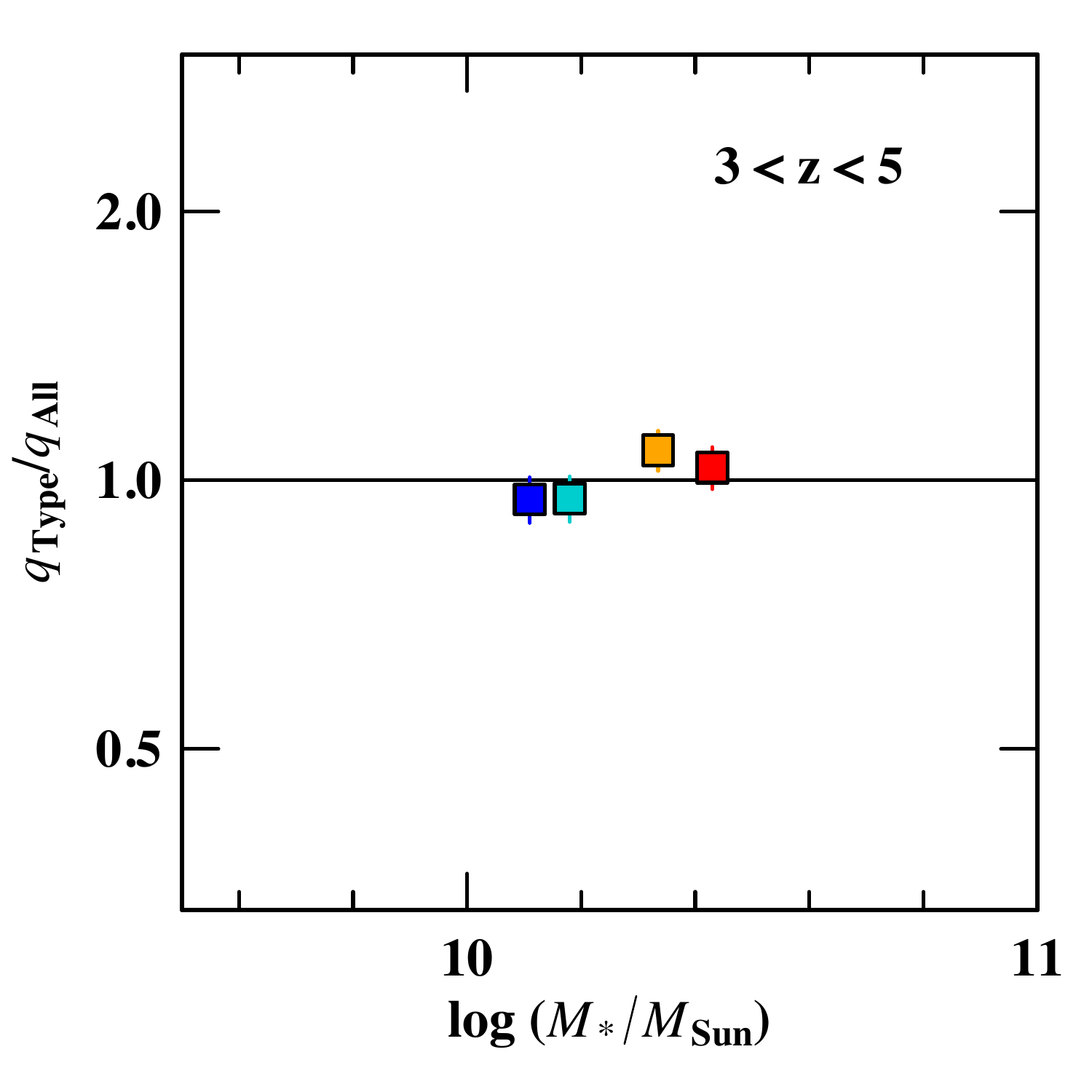}
\caption{Ratio of the (circularized) effective radius (left panel) and axis ratio (right panel) for different galaxy types to those of the parent SFG sample. Galaxies are in the redshift bin $3 < z < 5$ and are massive ($\log (M_{*}/M_{\odot}) > 10$). In the case of the effective radius, ratios calculated after applying a correction to a common redshift and stellar mass are displayed with open squares. The ratios are obtained by dividing the median values of a given galaxy type by the median values of the parent SFG sample, with the error bars representing the uncertainty of the ratio of the medians.}
\label{fig:ratio_morph_mstar}
\end{center}
\end{figure*}

Figure~\ref{fig:ratio_morph_mstar} shows the ratio between median (circularized) effective radius of the different galaxy types and the median (circularized) effective radius of the parent sample of SFGs. For better statistics, we focus on a single redshift bin ($3 < z < 5$) of massive galaxies ($\log (M_{*}/M_{\odot}) > 10$). In addition, we also move beyond the selection criteria for LBGs and OFGs by simply defining galaxies with higher dust attenuation ($A_V > 1$ sample), of which OFGs would be specific cases, and galaxies with lower dust attenuation ($A_V < 1$ sample), which would comprise the majority of LBGs. The $A_V > 1$ and OFG samples are indeed smaller than the parent sample of SFGs at the same redshift and stellar mass bin. In the case of axis ratios, these appear generally consistent with the values of the parent sample of SFGs, albeit slightly skewed toward larger values, which is consistent with the picture in which axis ratios are more difficult to constrain when the angular resolution starts to be comparable with the intrinsic size of the galaxy. This affects more galaxy types that are intrinsically smaller, such as the $A_V > 1$ and OFG samples and, generally, galaxies at higher redshifts following the decreasing sizes of SFGs with increasing redshift \citep[e.g.,][]{vanderwel14}. We note that the upper limits in the sizes for the 6/25 OFGs classified as point sources are not included here. If included, they would lower the size ratio even further, in line with $A_V > 1$ and OFG samples exhibiting smaller sizes than the parent sample of SFGs.

While we studied a given galaxy type compared with the parent sample of SFGs in the same redshift and stellar mass bin, there are still differences in the redshift and stellar mass distributions. In Fig.~\ref{fig:ratio_morph_mstar}, measurements (filled squares) are placed at the median stellar mass of the galaxy type. These galaxy types do not have the exact same median stellar mass (or redshift) between them and they do not have the exact same median stellar mass (or redshift) as the parent SFG sample. This could have an impact on the comparison. However, we know from literature studies that SFG sizes decrease with increasing redshift and increase with increasing stellar mass \citep[e.g.,][]{shen03,vanderwel14,barro17,suess19,mowla19}. While these studies were carried out in the pre-\textit{JWST} era, and therefore they are typically limited to $z < 3$ for stellar sizes in the rest-frame optical/near-IR, if we assume that the same trends can be extrapolated to higher redshifts, we can correct for the differential effect that redshift and stellar mass have on the sizes \citep[see also][for a similar analysis]{gomezguijarro22a}. Following this approach, we corrected the effective radii of each galaxy to a common redshift and stellar mass as given by the median values of the $3 < z < 5$ massive ($\log (M_{\rm{*med}}/M_{\odot}) > 10$) parent sample of SFGs (open squares in Fig.~\ref{fig:ratio_morph_mstar}). After this exercise, $A_V > 1$ and OFG samples are a factor 1.3 smaller than the parent sample of SFGs and a factor 1.5 smaller than the $A_V < 1$ sample at the same redshift and stellar mass bin.

\subsection{Random forest analysis} \label{subsec:rand_forest}

The results above indicate that stellar mass is a primary proxy for dust attenuation. In a general sense, effective radius and axis ratio do not play a clear role in dust attenuation. However, when considering massive galaxies at similar redshifts, highly dust-attenuated galaxies exhibit more compact (smaller effective radii) stellar light profiles. We check these results by performing a random forest analysis \citep[see e.g.,][for a similar analysis]{ellison20,baker22,bluck22}.

A random forest is a machine learning technique that identifies the most important parameter in driving a given quantity, especially when several parameters are inter-correlated. A target quantity is selected and removed from the data, leaving a dataset containing the features that the target quantity imprinted on it. These two datasets (target and features) are then split into a training and a test sample. The algorithm is employed in the training sample, which sorts the data in different nodes into the different decision trees with the goal being to minimize the Gini Impurity \citep{pedregosa11}. The final model is then applied to the test sample in order to check its performance.

We used a random forest regressor in order to identify the most predictive parameter of $A_V$ between $M_{*}$, SFR, $R_{\rm{e}}$, and $q$. These parameters are chosen to reflect different aspects of galaxies, going from the total amount of stars ($M_{*}$), the rate of creating new stars (SFR), the extent of the stellar light $R_{\rm{e}}$, and the way the stellar light is oriented ($q$). We also added a random uniform variable ($Rand$) as a control variable. Using a randomized cross-search validation method we fine-tuned the hyperparameters. We checked the performance of the fit by calculating the mean squared errors (MSEs) between model and test sample.

\begin{figure}
\begin{center}
\includegraphics[width=\columnwidth]{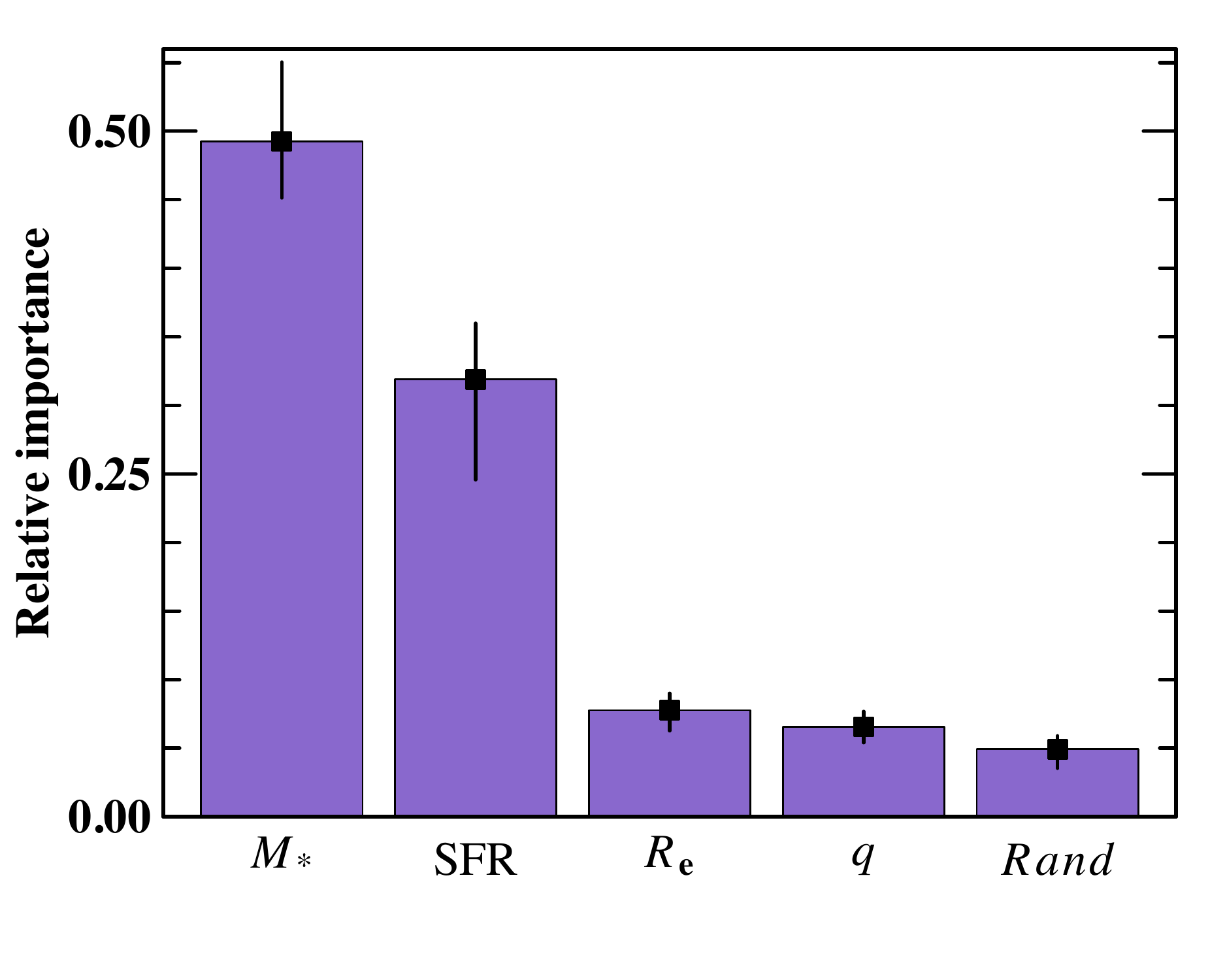}
\caption{Relative importance of the different parameters explored in predicting dust attenuation from a random forest regressor analysis. The bar height displays the median value of the distribution of 10\,000 realizations of training and test samples, while the error bars represent the 16\% and 84\% percentiles of the distributions.}
\label{fig:rf}
\end{center}
\end{figure}

Figure~\ref{fig:rf} shows the relative importance of the studied parameters in predicting $A_V$ from the random forest regressor. Stellar mass is the most predictive parameter, followed by SFR. The morphological parameters of effective radius and axis ratio do not appear as strong predictors of dust attenuation, with the axis ratio being close to random. These results are in line with the results shown in Sect.~\ref{sec:av_drivers}.

\section{Discussion} \label{sec:discussion}

The results in this work show that stellar mass is a primary predictor of dust attenuation. The correlation between dust attenuation and stellar mass was studied in the pre-\textit{JWST} era, especially at $z < 4$ using statistical galaxy samples \citep[e.g.,][]{garn10,zahid13,heinis14,pannella15,alvarezmarquez16}, with some studies peering into the $z > 4$ Universe using smaller samples \citep[e.g.,][]{fudamoto20,bogdanoska20}. In the \textit{JWST} era, some works started to investigate the relation between dust attenuation and stellar mass at $z > 3$ \citep[e.g.,][]{shapley23}, with models predicting similar trends \citep[e.g.,][]{yung19}. Stellar mass is a tracer of the integrated history of star formation in galaxies and therefore the history of metal and dust production. The stellar mass defines the depth of the galaxy potential well and therefore massive galaxies are able to keep their metals and dust more efficiently than low-mass galaxies, which are prone to lose them through galactic winds. The mass--metallicity relation is strong observational proof of these physical processes \citep[e.g.,][]{tremonti04,erb06,sanchez13,genzel15,ma15}.

Once  the stellar mass has been accounted for, SFR appears as another strong predictor of dust attenuation. The two are correlated in SFGs, as shown by the existence of the MS of SFGs \citep[e.g.,][]{brinchmann04,daddi07,elbaz07,noeske07,whitaker12,speagle14,schreiber15}. At fixed stellar mass, a number of studies in the literature explain the dust attenuation scatter through variations in the SFR \citep[e.g.,][]{garn10,wuyts11,zahid13}. However, one interesting aspect of the random forest analysis in Sect.~\ref{subsec:rand_forest} is that stellar mass appears as a stronger predictor of dust attenuation than SFR. While the mass--metallicity relation can be considered a first-order relation in galaxy evolution, the fundamental metallicity relation \citep[FMR;][]{mannucci10} comes as a second-order indicator of the metal content in galaxies. The FMR indicates that, at fixed stellar mass, galaxies with higher SFR have lower metallicities. Therefore, the FMR would favor stellar mass as a stronger predictor of dust attenuation than SFR.

Orientation has been studied in the literature as a proxy for dust attenuation. Disk galaxies can exhibit higher dust attenuation levels for higher inclinations, owing to the thicker projected dust columns \citep[e.g.,][]{wild11,patel12,zuckerman21}. \citet{zuckerman21} explains the dust attenuation scatter at fixed stellar mass as due to inclination effects in thin dust disks, but also shows that when the dust disk is thick, inclination becomes irrelevant. The results presented in this work show that at $z > 3,$ inclination has a negligible effect on dust attenuation, and is almost indistinguishable from a random variable. Highly dust attenuated systems in this work do not exhibit a preference for more inclined orientations (i.e., edge-on disks). In particular, OFGs are not biased toward edge-on disks, contrary to the interpretation of \citet{nelson23}, who links the obscuration of similar galaxies to an edge-on nature. A very recent study by \citet{lorenz23} also reaches the same conclusion at lower redshifts, showing that none of the dust properties studied (Balmer decrement, $A_V$, rest-frame UV slope) vary with axis ratio in a sample of 308 spectroscopically characterized SFGs at $1.3 < z < 2.6$. The authors propose an interpretation in which already at $z > 1$ the dust attenuation in SFGs would be dominated by dust in dense star-forming regions rather than by the dust in the diffuse ISM, which would mean a negligible effect of inclination on dust attenuation. Some examples of dust swept in dense clouds and depleted diffuse components have also been observed in the local Universe \citep{holwerda13,keel14}.

While stellar mass and SFR come as strong predictors of dust attenuation for the parent sample of SFGs, our results show that there is a subset of galaxies that, even at fixed stellar mass, SFR, and redshift, are fainter than the average SFG at rest-frame UV wavelengths with higher levels of dust attenuation ($A_V > 1$). OFGs are a specific case of these galaxies, and exhibit a factor 2.4 higher $A_V$ than the parent sample of SFGs (for $3 < z < 5$ and $\log (M_{*}/M_{\odot}) > 10$). We find that the key distinctive feature is their compact size (smaller effective radii) compared to the average SFG at the same stellar mass and epoch (as given by the parent sample of SFGs). The physical interpretation of this is that the galaxy size is the parameter that modulates how the density profile of the galaxy is distributed. More compact stellar profiles could therefore contribute to more efficiently keeping metals and dust in a galaxy compared to a more extended stellar profile. For galaxies with similar levels of dust mass (determined at first order by their redshift, stellar mass, and SFR), the dust column density would be thicker when embedded in more compact stellar profiles than in more extended stellar profiles.

More compact stellar profiles in OFGs are in line with the view of ALMA-detected OFGs; they exhibit enhanced SFR surface densities associated with compact dust morphologies with high dust covering fractions. Therefore, these galaxies are capable of forming stars in concentrated areas ---which would result in the compact stellar light profiles we see in this work--- and of obscuring the stellar light behind thick dust columns \citep[e.g.,][]{smail21,xiao23,kokorev23}. The compact stellar sizes of OFGs are comparable to those of QGs. Massive ($\log (M_{*}/M_{\odot}) > 10$) OFGs in the redshift bin $3 < z < 4$ have a median (circularized) effective radius of $R_{\rm{e}} = 0.75 \pm 0.25$\,kpc (where the uncertainty is given by the median absolute deviation). As a reference, QGs at $z = 3$ of similar stellar mass have a typical effective radius of $R_{\rm{e}} = 0.73$\,kpc \citep{vanderwel14}. In the classical view of galaxy structural evolution, SFGs evolve mostly as extended star-forming disks, while QGs are typically more compact than SFGs at a fixed stellar mass and redshift \citep[e.g.,][]{shen03,vanderwel14,barro17,suess19,mowla19}. The buildup of a central stellar core appears to be linked to the quenching of star formation \citep[e.g.,][]{kauffmann03,whitaker17,barro17,gomezguijarro19,suess21}. SFGs with compact star forming regions \citep[e.g.,][]{toft14,gomezguijarro18,gomezguijarro22b,franco20,puglisi21,magnelli23} and SFGs with a developed compact stellar core \citep[e.g.,][]{barro13,nelson14,williams14,vandokkum15,gomezguijarro19} have been proposed as the link between the more extended SFGs and the more compact QGs. The compact sizes (smaller effective radius) in OFGs when compared to the general SFG population shown in this work and the higher SFR surface densities in OFGs shown in literature studies are in line with a progressive buildup of the compact  stellar profiles of QGs. Therefore, OFGs (at a higher redshift bin) could be progenitors of massive compact QGs (at a lower redshift bin).

\section{Summary and conclusions} \label{sec:summary}

In this work, we investigate the drivers of dust attenuation in the \textit{JWST} era, with a special focus on understanding why the galaxies known as optically faint/dark galaxies (OFGs) are so faint at optical/near-IR wavelengths. To this end, we use data from \textit{JWST} to unveil their nature and characterize their basic stellar and morphological properties. Our findings can be summarized as follows:

\begin{itemize}

\item Stellar mass is a primary proxy of dust attenuation, among the properties studied. Effective radius and axis ratio do not show a clear link with dust attenuation, with the effect of orientation being found to be close to random.

\item There is a subset of highly dust attenuated ($A_V > 1$) SFGs, of which OFGs are a specific case. Even at the same stellar mass, SFR, and redshift, this subset of galaxies are still fainter than the average SFG at rest-frame UV wavelengths with high levels of dust attenuation. We find that their key distinctive feature is their compact size, with 30\% smaller effective radius than the average SFG at the same stellar mass and redshift (for massive systems with $\log (M_{*}/M_{\odot}) > 10$). These galaxies do not show a preference for low axis ratios (i.e., edge-on disks).

\item The OFG selection criteria ($F150W > 26.5$\,mag; $F444W < 25$\,mag) are an effective filter for relatively high-redshift ($3 < z < 7.5$), intermediate to massive ($\log (M_{*}/M_{\odot}) \geq 9.0$), dusty ($A_V \gtrsim 1$) SFGs, with little contamination from QGs ($\sim 8.6$\%). Compared to the general SFG population, OFGs exhibit smaller effective radii, with S\'ersic indices of $n \sim 1$ and axis ratios of around $q \sim 0.5$.

\end{itemize}

The results in this work show that stellar mass is a primary proxy for dust attenuation and compact stellar light profiles behind the thick dust columns obscuring typical massive SFGs.

\begin{acknowledgements}

CGG acknowledges support from CNES. PGP-G acknowledges support  from Spanish  Ministerio  de  Ciencia e Innovaci\'on MCIN/AEI/10.13039/501100011033 through grant PGC2018-093499-B-I00. This work is based on observations made with the NASA/ESA/CSA James Webb Space Telescope. The data were obtained from the Mikulski Archive for Space Telescopes at the Space Telescope Science Institute, which is operated by the Association of Universities for Research in Astronomy, Inc., under NASA contract NAS 5-03127 for JWST. These observations are associated with program \#1345. We would acknowledge the following open source software used in the analysis: \texttt{Astropy} \citep{astropy1,astropy2,astropy3}, \texttt{photutils} \citep{photutils}, \texttt{APLpy} \citep{aplpy}, \texttt{pandas} \citep{pandas}, \texttt{NumPy} \citep{numpy}. We are grateful to the anonymous referee, whose comments have been very useful to improving our work.

\end{acknowledgements}

\bibliographystyle{aa}
\bibliography{jwstceers_obs.bib}

\end{document}